# Persistent Spread Measurement for Big Network Data Based on Register Intersection

You Zhou
Department of CISE
University of Florida
youzhou@cise.ufl.edu

Yian Zhou
University of Florida
& Google Inc.
yian@cise.ufl.edu

Min Chen
University of Florida
& Google Inc.
min@cise.ufl.edu

Shigang Chen
Department of CISE
University of Florida
sgchen@cise.ufl.edu

## ABSTRACT

Persistent spread measurement is to count the number of distinct elements that persist in each network flow for predefined time periods. It has many practical applications, including detecting long-term stealthy network activities in the background of normal-user activities, such as stealthy DDoS attack, stealthy network scan, or faked network trend, which cannot be detected by traditional flow cardinality measurement. With big network data, one challenge is to measure the persistent spreads of a massive number of flows without incurring too much memory overhead as such measurement may be performed at the line speed by network processors with fast but small on-chip memory. We propose a highly compact Virtual Intersection HyperLogLog (VI-HLL) architecture for this purpose. It achieves far better memory efficiency than the best prior work of V-Bitmap, and in the meantime drastically extends the measurement range. Theoretical analysis and extensive experiments demonstrate that VI-HLL provides good measurement accuracy even in very tight memory space of less than 1 bit per flow.

## 1. INTRODUCTION

Massive and distributed data are increasingly prevalent in modern networks as high-speed routers forward packets at hundreds of gigabits or even terabits per second. Big data also happens at the network edge. For a few examples, Google handles over 40,000 search queries every second [1], and 500 million tweets are produced per day [2]. Traffic measurement and classification at such high speeds and with such massive volumes pose significant challenges [3, 4, 5, 6, 7, 8, 9, 10, 11, 12, 13, 14]. Exact measurement of big network data is often infeasible due to excessively high memory requirement and computation/communication overhead, whereas approximate estimation with probabilistic guarantees is a viable option.

Flow cardinality estimation [15, 16, 17, 18, 19, 20] is a fundamental problem in network traffic measurement. It estimates the number of *distinct* elements in every flow during pre-defined measurement periods. Each *flow* is uniquely identified by one or multiple fields in the packet headers, called *flow label*, which can be flexibly defined based on application needs. As examples, the flows under measurement may be per-source flows (with flow label being the source address), per-destination flows, TCP flows, WWW flows, or application-specific flows. The *elements* under measurement can be destination addresses, source addresses, ports, values in other header fields, or even keywords that appear in packet payload. For example, for each per-source flow, if destination addresses are treated as elements, then a flow's cardinality is the number of distinct destination addresses that the flow source has contacted, which can be used for scan detection.

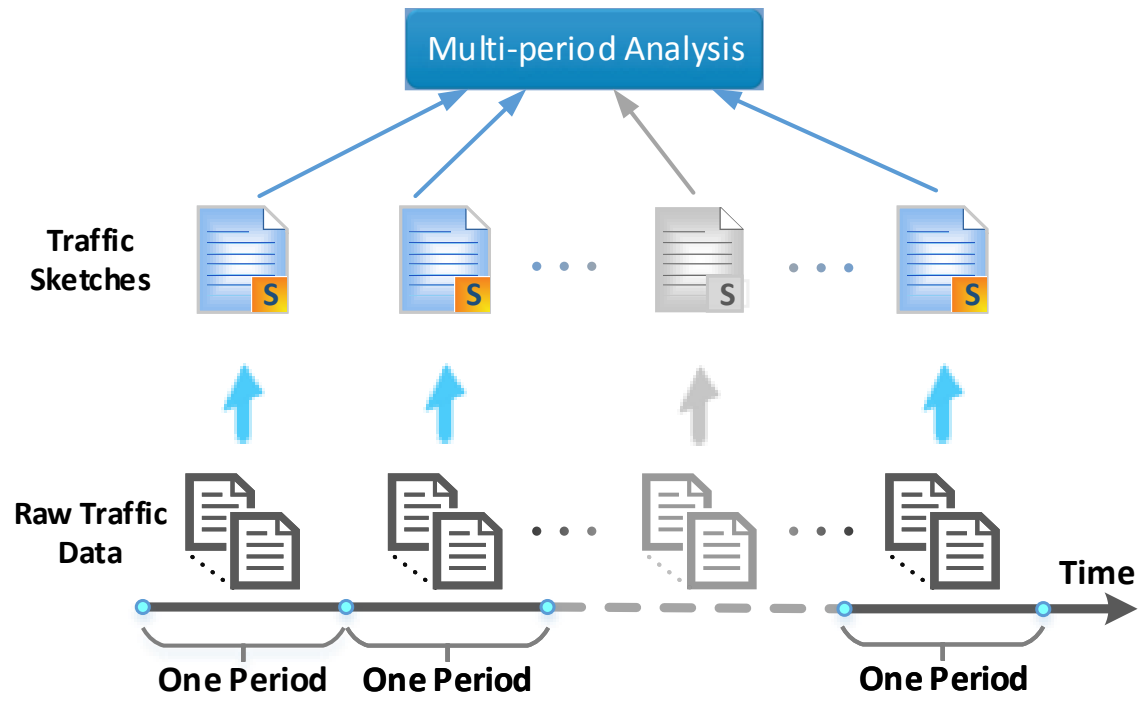

Figure 1: Multi-period analysis of data sketches.

Existing research on flow cardinality estimation mainly focuses on analysing traffic sketches from one measurement period, which is the summary of the raw traffic data in that time period. Since online storage can only hold limited information, the sketches are usually offloaded to a server after each measurement period for long-term storage and offline query. This paper studies an under-investigated problem of analyzing sketches across multiple periods as shown in Figure 1. In particular, we are interested in measuring the *persistent spread* of each flow, which is defined as the number of distinct elements that show up in a network flow during a certain number of consecutive measurement periods.

**Practical Importance:** Persistent spread measurement has many practical applications. Traditional super-spreader detection is to identify the "elephant" flows whose cardinalities are abnormally large, and can be applied to monitoring network anomalies. For instance, scanners may be identified if they send probes to too many destination addresses, i.e., the cardinalities of per-source flows are large. But there are practical scenarios where flow cardinality alone is in-

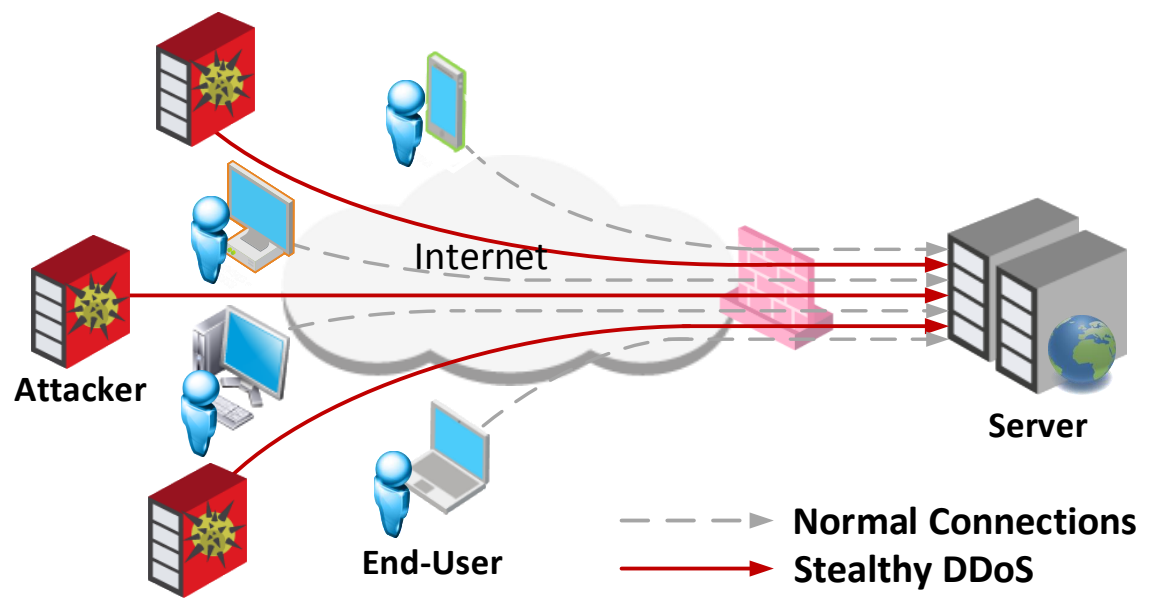

Figure 2: Stealthy DDoS attack.

adequate — a stealthy scanner may intentionally reduce its probing rate to reduce its flow cardinality in order to evade detection. Even with a reduced probing rate, after sufficient time, the scanner can still discover systems with vulnerability to exploit. In this case, measuring persistent spread can help identify such stealthy scanners. As a scanner probes different destination addresses over time, its persistent spread is zero or low; if a scanner deliberately repeated many of the same destinations, it would significantly slow down the already small scanning rate. Therefore, modest flow cardinality but usually low persistent spread helps signal a low-rate scanner that wanders in the destination address space. In the second example, DDoS attacks may be identified if unusually many clients send requests to a server, i.e., the cardinality of a per-destination flow is too high. However, as illustrated in Figure 2, with a smaller number of available attacking machines, a stealthy DDoS attack does not attempt to overwhelm the target server with excessive requests, but to degrade its performance [21]. If the number of attacking machines is similar to the number of legitimate users, we will not observe unusual flow cardinalities. Again, measuring persistent flow cardinality may help. According to the study [22] of real-world network traces from CAIDA [23], the continuous interaction between legitimate users and their target servers is normally shorter than twenty minutes. For attackers, since their goal is to degrade the performance of the target server over a long period, these hostile machines will send requests persistently to the target server, resulting in a significant persistent cardinality over time that is higher than usual.

Persistent spread measurement also has applications at the network edge (e.g., web search and social media). Take Google trends as an example. If Google treats all client IPs that query a keyword as a flow, the cardinality of the flow suggests the popularity of the keyword being searched. However, a significant number of colluding machines with different IP addresses can periodically query the same keywords, and make these keywords popular in Google trends as they wish. Since normal users typically do not query the same keywords periodically for a long time, persistent spread measurement can help detect such long-term search patterns, where a large set of IPs keep querying the same keywords over multiple periods. Besides detecting faked popularity, our work may serve as a generalized primitive tool for detecting hidden activities that manifest only over long time.

**Prior Art and Challenges:** Most previous work focuses on traffic sketches of one measurement period. To deal with a large number of flows, a series of sketches were developed to reduce massive raw data to a summary of per-flow cardinalities during online measurement. These solutions include PCSA [16], Multi-Resolution Bitmap [15], LogLog [17], and HyperLogLog (HLL) [18]. The principle is to allocate a separate data structure, containing a certain number of bitmaps, registers or other elementary data structures, to each flow for recording its elements. Over the past decades, a major research thrust is to reduce the sketches' memory footprint. But it has been a difficult undertaking with slow progress. For instance, per-flow memory requirement for cardinality measurement was reduced from thousands of bits to hundreds of bits by HLL [18], which ensures a large measurement range with good accuracy.

However, as the Internet enters the big-data era, hundreds of bits per flow can still be too much when there are too many flows. An example is modern high-speed routers, which forward packets from incoming ports to outgoing ports via switching fabric at the extraordinary speeds. To sustain high throughput, online modules for packet scheduling, access control, quality of service and traffic measurement are often implemented on network processors, bypassing main memory and CPU almost entirely. The on-die memory (such as SRAM) in a network processor is fast but small, and may have to be shared by multiple functions. Therefore, it is highly desirable to implement these functions as compact as possible. As this paper focuses on persistent cardinality measurement, we want to push its memory usage to an unprecedented low level, in order to save space for other functions on the same chip.

In another example, suppose a web-search analyst wants to profile, for each keyword (phrase, question or sentence), the number of distinct users that have searched the keyword. This information is useful in online social/economical/opinion trend studies or optimizing search performance [24]. As we have discussed earlier, persistent spread measurement can be used to detect faked popularity. However, since the number of flows (one flow per keyword, phrase, question or sentence) can be in many billions, it presents a challenge in computational resources, and memory in particular. Instead of using an expensive and powerful server, if we can drastically reduce the resource requirement, we may be able to run such analysis on a cheap commodity computer, which is a welcome result when high-end machines are not readily available.

To sum up, there are practical scenarios with great disparity between memory demand and availability, which requires online cardinality measurement to be implemented as compact as possible. Moreover, the design of a measurement function should also ensure reasonable accuracy with a large measurement range that supports "elephant" flows with very high persistent cardinalities.

To the best of our knowledge, little research work on persistent spread measurement exists in literature. Chen et al. [5] propose a continuous variant of Flajolet-Martin sketches adapted from [16], which however cannot give accurate results when the available memory space is tight [22]. Xiao et al. [22] design a bit sharing architecture called multi-virtual bitmaps, which store a flow's information in a virtual bitmap during each measurement period and analyzes the bitmaps from multiple periods to find persistent cardinality. The major drawback is that the measurement range of bitmaps is very small and no more than a few thousands for a typical implementation.

**Our Contributions:** The objective of our research is to improve the memory efficiency and enlarge the range of persistent spread measurement, while keeping good accuracy. Our main contributions are summarized below.

First, we design a highly efficient persistent spread estimator called Intersection HLL (I-HLL) that works over multiple measurement periods. Every flow is allocated a separate HLL sketch of registers to record its cardinality in a measurement period. We apply register intersection over the series of HLL sketches produced for a flow during a given number of measurement periods. We then employ maximum likelihood estimation to develop the formula of the I-HLL estimator that computes an estimate of the flow's persistent spread. We formally analyze the accuracy of the estimation, and show I-HLL has a large estimation range.

Second, to further improve memory efficiency, we introduce register sharing on top of I-HLL and propose a highly compact Virtual Intersection HLL (VI-HLL) architecture to measure the persistent spreads of a large number of flows simultaneously. Similar to [20], each flow is allocated a virtual HLL sketch of multiple registers, and the virtual HLL sketches of all flows share a common pool of phys-ical registers. But unlike [20] that measures flow cardinality in one period, our VI-HLL deals with persistent cardinality over multiple periods. VI-HLL achieves far better memory efficiency and much larger range than the best existing work (V-Bitmap [22]) on persistent cardinality measurement.

Finally, not only do we mathematically analyze the estimation accuracy of VI-HLL, but also perform extensive experiments to compare it with V-Bitmap. The experimental results demonstrate the superior performance of VI-HLL. Interestingly, its estimation accuracy improves when the number of measurement periods increases.

# 2. PROBLEM STATEMENT

Consider the packet stream arriving at a router (or firewall) inside a high-speed network or the application records produced by a server (e.g., web search) at the network edge. We model both types of network data as a sequence of $\langle$flow label, element$\rangle$ pairs in our abstraction. Based on the flow labels, the sequence of pairs are classified into different flows. For the packet stream as example, if we want to measure the number of distinct sources that have contacted each destination, we abstract every packet as a pair of destination address and source address, which can both be extracted from the packet header. All pairs (i.e., packets) with the same destination address (i.e., flow label) constitute a flow. In the example of web search, each search record is abstracted as a pair of keyword and source address (from which the search request is received). All pairs with the same keyword are treated as a flow.

We are interested in measuring elements that keep showing up over time in each flow. The issue is how to quantitatively define the persistency of "keep showing up over time". Consider the traditional definition of flow cardinality (or spread) measurement [15, 16, 17, 18], which is to find the number of distinct elements in each flow during a certain time frame $[0, T]$. This definition does not capture the property of persistency. We illustrate it through an example of measuring the number of distinct sources that have contacted each destination, where all packets to the same destination form a per-destination flow. Suppose one million different sources contacted a destination during a day. The cardinality of this per-destination flow is one million. But if all the sources contacted the destination in the first 10 minutes and no contact was made for the rest of the day, we cannot say these sources "kept contacting" the destination for the day. The persistent spread is zero in this case. To formulating persistency, one way is to divide the day into measurement periods of 10 minutes each. If we find that 1000 sources out of the million were present in each period, they were the persistent elements that we want to measure. The remaining elements that showed up only in the first period were not persistent. So the persistent spread is 1000.

Generally, we specify persistency by dividing time into measurement periods and measure those elements that are present persistently in a pre-set number $t$ of consecutive periods under consideration. We give a more formal definition as follows: Consider an arbitrary flow and $t$ consecutive measurement periods. Let $S_j$ be the set of distinct elements in the flow observed during the $j$th measurement period, $1 \leq j \leq t$. Let $S^*$ be the subset of common elements observed in all $t$ periods, i.e., $S^* = S_1 \cap S_2 \cap \ldots \cap S_t$. The problem of *persistent spread measurement* is to find the size of $S^*$, denoted as $n^* = |S^*|$, which is called the *persistent spread* of the flow. The elements in $S^*$ are called *persistent elements*. The elements in $S_j - S^*$, $1 \leq j \leq t$, are called *transient elements*.

The proposed architecture for estimating the flows' persistent spreads is intended to be generic, while its parameters should be set by system admins based on their application needs. In particular, the length of each period and the number $t$ of periods used are application-dependent. As an analogy, a system admin will configure the threshold for scan detection (i.e., the triggering number of different destinations that a source contacts over a period) to be more than the

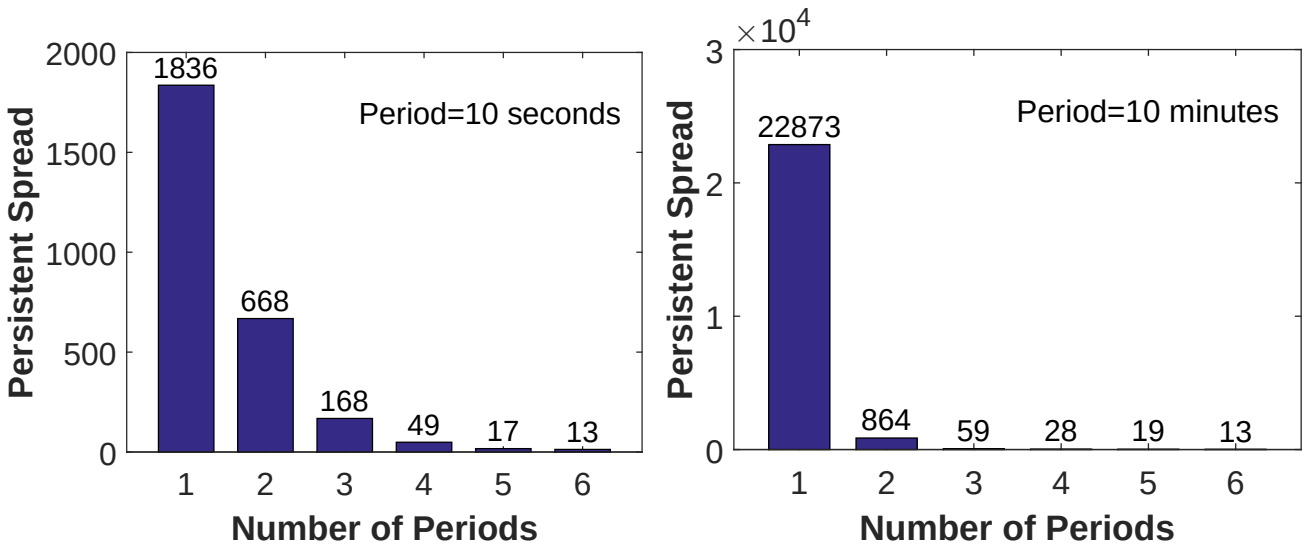

Figure 3: Persistent spread of a packet flow to destination 97.208.145.236, with respect to different period lengths in the two plots and different numbers $t$ of periods on the horizontal axis.

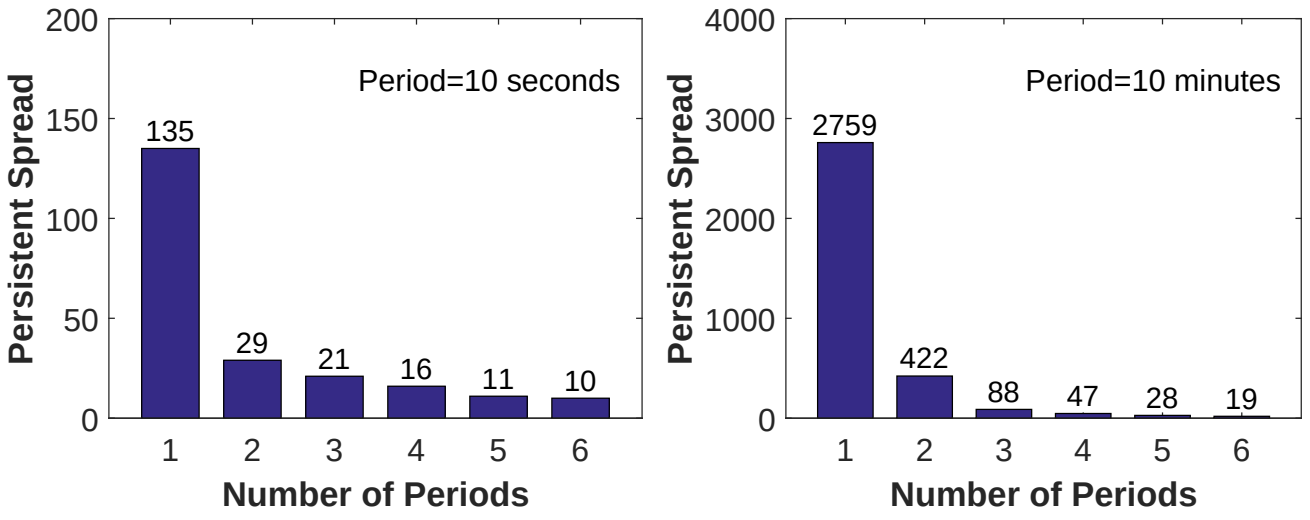

Figure 4: Persistent spread of a packet flow to destination 220.221.80.140, with respect to different period lengths in the two plots and different numbers $t$ of periods on the horizontal axis.

*measured numbers* of most normal sources, which may vary from network to network. Similarly, the parameters of persistent spread measurement should also be set based on application-specific and system-specific normal traffic statistics. Consider the example in the introduction on detecting stealthy DDoS attacks by measuring persistent spreads of per-destination (server) flows. If we set the measurement period to be a day, we may find significant persistent spreads for servers in normal traffic, because legitimate users may regularly access their email, web and other services on a daily basis. If we set the measurement period to be a few seconds, we may still find significant persistent spreads in normal traffic because any single connection to a service may last for many consecutive periods. However, if we choose a period length in-between and use a sufficient number of periods, it becomes unlikely for many normal users to exhibit the same persistency in accessing the servers as the attacking hosts [22].

The above analysis is confirmed by our experiments using a real network traffic trace from CAIDA, containing 39,456 per-destination flows in an hour. We vary the length of measurement period and the number $t$ of periods when measuring the persistent spreads of the flows. The measurement results for two randomly-selected large flows are shown in Fig. 3-4, and the statistics of all flows are shown in Fig. 5. Consider the flow in Fig. 3. Both plots show that its persistent spread drops quickly when we increase the number of periods. However, in the left plot where the period is short (10 seconds), if the number $t$ of periods used is too small (e.g., 2 or 3), the persistent spread of this normal traffic can be significant. For example, when $t = 2$, the persistent spread is 668, which is 36% of the spread when $t = 1$, i.e., the number of active sources in one period. Similar observation can be made in Fig. 4. On the other hand, as we increase the period length to 10 minutes in the right plot, when $t = 2$, the persistent spread is just 3.8% of the spread when $t = 1$. Be aware that a period of 10 minutes has many more packets (thus elements) than a period of 10 seconds; therefore, the relative percentage (36% v.s. 3.8%) is a better indicator for the impact of period length on persistent spread. By choosing a period length of 10 minutes and letting $t = 6$, we observe just 13 persistent elements (sources) in the flow during an hour. In contrast, when the period length is 10 seconds and $t = 6$, we observe 198 persistent elements in an hour. Fig. 5 presents the flow distribution with respect to the spread (or cardinality) value when $t = 1, 2, 3$ and 6 in the four plots, respectively. We put flows in bins with spread ranges of [0, 10], (10, 50], (50, 100], ... The length of each period is 10 minutes. The figure shows that most flows in this normal traffic trace have small spreads. When we increase the number of periods, the number of flows in bins of large spreads decreases quickly, suggesting that the persistent spreads of those flows are reduced to small values. This property helps in anomaly detection: When we see the persistent spread of a per-destination flow suddenly jumps from a usually small value to a large one, it signals a possible DDoS attack as we explain earlier in the introduction.

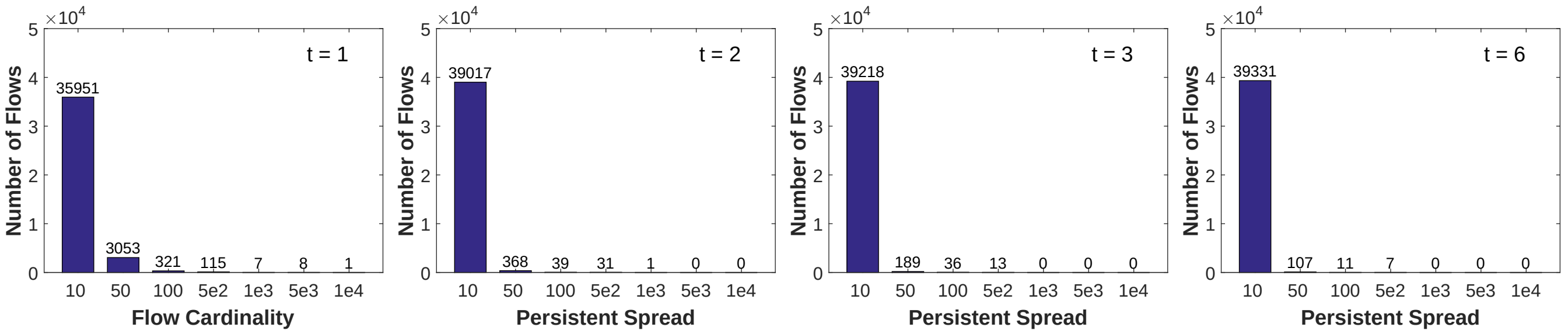

Figure 5: Flow distribution with respect to persistent spread under different $t$ values.

The objective of this paper is to design a persistent spread estimation architecture that consists of an online component and an offline component, where the former records all elements from all flows in real time using highly-compact data structures — which keep only sketches of the raw traffic data and are offloaded to a server after each measurement period, and the latter performs persistent spread estimation based on the sketches from multiple periods. We will evaluate the performance of our design based on the following two metrics.

*Memory overhead:* The disparity in memory demand and supply for practical traffic measurement scenarios explained in the introduction motivates us to make the online component of persistent spread measurement as compact as possible.

*Estimation accuracy:* Let $\hat{n}^*$ be the estimation result of the actual persistent spread $n^*$ of a flow. The estimation accuracy is evaluated based on the relative bias, $Bias(\frac{\hat{n}^*}{n^*})$, and the relative standard error, $StdErr(\frac{\hat{n}^*}{n^*})$, which are defined below.

$$\begin{aligned} Bias(\frac{\hat{n}^*}{n^*}) &= E(\frac{\hat{n}^*}{n^*}) - 1, \\ StdErr(\frac{\hat{n}^*}{n^*}) &= \sqrt{Var(\frac{\hat{n}^*}{n^*})} = \frac{\sqrt{Var(\hat{n}^*)}}{n^*}. \end{aligned} \tag{1}$$

Clearly, smaller values of relative bias and relative standard error mean more accurate measurement results. Given a certain available memory space, we want to make persistent-spread estimations as accurate as possible.

We make two assumptions, which are needed by our statistical analysis. The first assumption is that there are a large number of flows in each period and the number of distinct elements/persistent elements in any flow is negligibly small when comparing with the total number of distinct elements/persistent elements in all flows. The second assumption is that transient elements can be approximately treated as being independent among different periods. The same assumptions are needed in [22], which provides network traffic analysis to support the assumptions. The observation is that when the length of each period is sufficiently long, most transient elements will stay in one period because most user connections do not take that long. For example, when the period is set to 7 minutes with a gap of 3 minutes between consecutive periods, traffic analysis in [22] shows that less than 5% of all HTTP connections overlap with more than one period. The percentage will be lower if the period is set longer.

# 3. PRELIMINARIES

In this section, we first introduce the HyperLogLog (HLL) algorithm [18], and then present a straightforward register-union approach for persistent spread estimation based on HLL, which further motivates a more accurate register-intersection approach.

## 3.1 HyperLogLog (HLL) Algorithm

The HLL algorithm has made impact on IT industry [19]. It is designed to estimate the number of distinct elements in a single stream (flow) during a single measurement period. HLL ensures a large estimation range and a good estimation accuracy. An incoming stream is modeled as a multi-set $S$, whose elements are in the domain $\mathcal{D}$. An HLL sketch $M$ of $s$ registers are allocated to store the cardinality information. Without loss of generality, let $s = 2^b, b \in \mathbb{N}$. The $i$th register in $M$ is denoted by $M[i]$, $i \in [0, s)$. The size of registers is set based on the maximum range of the cardinalities to be estimated. Specifically, a register with 5 bits can measure cardinalities up to $2^{2^5} \approx 4 \times 10^9$.

Algorithm 1 summarizes how to generate an HLL sketch for stream $S$. First, we initialize all $M[i]$ to zeros, $i \in [0, s)$. Let $h : [\mathcal{D}] \rightarrow [0, 1] \equiv \{0, 1\}^L$ be a suitable hash function that maps an element in domain $\mathcal{D}$ uniformly at random to the binary range of $L$ bits long. Let $\rho(q)$ be the position of the leftmost 1 for a binary string $q \in \{0, 1\}^L$, i.e., it equals one plus the length of leading zeros in $q$. For example, if $q = \langle 0001\ldots \rangle$, then $\rho(q) = 4$. For an incoming element $e$ in stream $S$, let $x$ be the binary representation of hash value $h(e)$, where $p$ is the leading $b$ bits in $x$, and $q$ is the remaining bits. Then the element $e$ is mapped to $M[p]$, and $M[p]$ is updated by

$$M[p] := \max(M[p], \rho(q)). \tag{2}$$

In other words, the stream $S$ is split into $s$ substreams, each of which is encoded in a register based on the first $b$ bits of hashed value $h(e)$. Each register is set to the maximum value of $\rho(q)$ among all elements $e$ in the corresponding substream. If no element is encoded by a register, the register remains zero.

At the end of a period, HLL estimates the number of distinct elements encoded by its sketch $M = \{M[0], M[1], \ldots, M[s-1]\}$ through normalized harmonic mean [18]:

$$\hat{n}_S = \alpha_s \cdot s^2 \cdot \Big(\sum_{i=0}^{s-1} 2^{-M[i]}\Big)^{-1}, \tag{3}$$

**Algorithm 1** HLL Sketch for a stream $S$

1: Initialize a register array $M$ of size $s = 2^b$ with all zeros;
2: **for** $e \in S$ **do**
3: $\quad x := h(e); \quad p := \langle x_1 x_2 \ldots x_b \rangle; \quad q := \langle x_{b+1} x_{b+2} \ldots \rangle;$
4: $\quad M[p] := \max(M[p], \rho(q));$
5: **end for**
6: **return** $M$ at the end of a measurement period

where $\alpha_s$ is the bias correction constant that is

$$\alpha_s = \left( s \int_0^\infty \left( \log_2 \left( \frac{2+u}{1+u} \right) \right)^s du \right)^{-1}. \tag{4}$$

Pre-computed values of $\alpha_s$ may be used in practice: $\alpha_{16} = 0.673$, $\alpha_{32} = 0.697$, $\alpha_{64} = 0.709$, and $\alpha_s = 0.7213/(1 + 1.079/s)$ for $s \geq 128$. According to [18], the estimation standard error is

$$StdErr\Big(\frac{\hat{n}_S}{n_S}\Big) = O\Big(\frac{1}{\sqrt{s}}\Big). \tag{5}$$

It has been shown that estimation by (3) is severely biased when the cardinality is smaller than $2.5s$. Hence, when the estimated cardinality from (3) is smaller than $2.5s$, we treat $M$ as a bitmap of $s$ bits, with each register $M[i]$ converted to one bit, whose value is one when $M[i] > 0$ or zero otherwise. The estimation formula for small cardinality is

$$\hat{n}_S = -s \cdot \ln V, \tag{6}$$

where $V$ is the fraction of bits in the bitmap whose values remain zeros.

## 3.2 HLL-Based Persistent Spread Estimation

The HLL sketches can be adopted for persistent spread estimation. To make technical discussion more concrete, we consider per-destination flows passing a router and measure the number of distinct source addresses in each flow. For an arbitrary flow, we allocate an HLL sketch $M$ of $s$ registers to record the flow's source addresses in each period. Denote the HLL sketch of the $j$th period by $M_j$.

At the beginning of the $j$th period, all registers of HLL sketch $M_j$ are initialized to zeros. When the router receives a packet, it extracts the flow label (i.e., destination address $dst$) from the packet header, and records the element (i.e., source address $src$) in $M_j$ by Algorithm 1. By the end of the period, the router has recorded the set $S_j$ of elements in $M_j$. It offloads $M_j$ to a server for long-term storage and offline query.

After $t$ consecutive periods, we have a sequence of HLL sketches $M_1$, $M_2$, ..., $M_t$. The problem is how to use these HLL sketches to estimate the persistent spread $n^* = |S^*| = |S_1 \cap S_2 \cap \ldots \cap S_t|$, which is the number of distinct elements that are present persistently through the $t$ periods. We propose two approaches, register union and register intersection, to solve this problem.

### 3.2.1 Register-Union Approach

According to the inclusion-exclusion rule, the cardinality of an arbitrary set intersection, including $n^*$, can be expressed as sums/differences of the cardinalities of set unions. The cardinality of any set union can be estimated using the HLL estimator (3) after performing register-wise union on the corresponding sketches. For example, the cardinality of set intersection $S_1 \cap S_2$ is

$$|S_1 \cap S_2| = |S_1| + |S_2| - |S_1 \cup S_2|. \tag{7}$$

Namely, $|S_1 \cap S_2|$ is represented as the sum/difference of three cardinalities, $|S_1|$, $|S_2|$ and $|S_1 \cup S_2|$, where $|S_1|$ and $|S_2|$ can be estimated from $M_1$ and $M_2$ using the HLL estimator, respectively. Moreover, given the sketches $M_1$ and $M_2$ for $S_1$ and $S_2$, the HLL sketch for the set union $S_1 \cup S_2$ is simply register-wise union $M_\cup = M_1 \vee M2$, where operator $\vee$ is defined to be $M_\cup[i] = \max(M_1[i], M_2[i]), 0 \leq i < s$. After that, $|S_1 \cap S_2|$ can be estimated by applying the HLL estimator on $M_\cup$. Generalizing the above analysis to intersection over more than two periods (i.e., $t > 2$) is straightforward.

Despite its mathematical simplicity, register-union estimate is very inaccurate since it does not fully explore the correlation among the $t$ HLL sketches. Let $n_\cup$ be the cardinality of set union $S_1 \cup S_2 \cup \ldots \cup S_t$, $n^*$ be the cardinality of set intersection $S^*$, and $\hat{n}^*$ be its estimate using the register-union approach. According to [5, 8], the estimation standard error of $n^*$ is

$$StdErr\Big(\frac{\hat{n}^*}{n^*}\Big) = O\Big(\frac{n_\cup}{\sqrt{s}n^*}\Big). \tag{8}$$

The estimation accuracy depends on $\frac{n_\cup}{n^*}$, and $StdErr\big(\frac{\hat{n}^*}{n^*}\big)$ increases as $\frac{n_\cup}{n^*}$ becomes larger. When $t$ is set large, $n_\cup$ may become large due to addition of more transient elements, whereas $n^*$ may stay more or less the same if the set of persistent elements does not change much, which drives up $\frac{n_\cup}{n^*}$ and thus inaccuracy in estimation. The accuracy loss as $t$ grows can prohibit a network admin from configuring a large value for $t$.

### 3.2.2 Register-Intersection Approach

By contrast, the register-intersection approach calculates the intersection of HLL sketches, $M_\cap = M_1 \wedge M_2 \wedge \ldots \wedge M_t$, where operator $\wedge$ on two arbitrary HLL sketches is defined as $(M_{j_1} \wedge M_{j_2})[i] = \min(M_{j_1}[i], M_{j_2}[i]), 0 \leq i < s$. Therefore, the value of the $i$th register in the intersection sketch $M_\cap$ is the minimal value of all corresponding registers in the $t$ original HLL sketches,

$$M_\cap[i] = \min_{j \in [1,t]} \big\{ M_j[i] \big\}, \quad 0 \leq i < s. \tag{9}$$

Let $M^*$ be the imaginary HLL sketch $M^*$ that records the true intersection set $S^*$. We derive the relationship between $M_\cap$ and $M^*$. As stated earlier, each flow element pseudo-randomly picks a register in $M_j$ by the hash function $h$, and updates the chosen register accordingly. Moreover, in different periods, a persistent element $e$ in $S^*$ always tries to update the same register $M_j[p]$ using the same value $\rho(q)$ as suggested by (2). Therefore, the same sketch $M^*$ constructed with the persistent elements from $S^*$ is embedded in $M_j$, $1 \leq j \leq t$, which can be considered as $M^* \vee T_j$, where $T_j$ is the sketch constructed with the transient elements, $S'_j = S_j - S^*$, in the $j$th period. $T_j$ changes from period to period as transient elements vary over time. Hence, we have

$$\begin{aligned} M_\cap &= M_1 \wedge M_2 \wedge \ldots \wedge M_t \\ &= (M^* \vee T_1) \wedge (M^* \vee T_2) \wedge \ldots \wedge (M^* \vee T_t) \\ &= M^* \vee (T_1 \wedge T_2 \wedge \ldots \wedge T_t). \end{aligned} \tag{10}$$

The value of the $i$th register in $M_\cap$ is, for $0 \leq i < s$,

$$M_\cap[i] = \max\{M^*[i], \min\{T_1[i], T_2[i], \ldots, T_t[i]\}\}. \tag{11}$$

If we use $M_\cap$ to approximate $M^*$, the HLL estimator will produce a result with positive bias because it may happen that transient elements set the $i$th registers in all $t$ periods higher than $M^*[i]$, causing overestimation of the persistent spread, although the probability for this to happen decreases as the number $t$ of periods increases. We will address this overestimation issue by maximum likelihood estimation.

# 4. INTERSECTION HLL ESTIMATOR

In this section, we present an Intersection HLL (I-HLL) estimator based on register intersection $M_\cap$ and maximum likelihood estimation (MLE) to measure the persistent cardinality of any flow.

## 4.1 Probability of Intersection Register Value

We first analyze the probabilistic distribution for an arbitrary register in the I-HLL sketch $M_\cap$, which will be used to construct an MLE estimator. Suppose we measure a flow over $t$ periods, and obtain a sequence of HLL sketches, $M_1$, $M_2$,. . ., $M_t$, which record element sets, $S_1$, $S_2$, . . ., $S_t$, respectively. Let $n_j$ be the cardinality of $S_j$. The number of elements in the transient subset $S'_j$ is $n'_j = n_j - n^*$.

We know $M_j = M^* \vee T_j$, i.e., the HLL sketch $M_j$ of the $j$th period is the combination of $M^*$ for the persistent elements and $T_j$ for the transient elements. In other words, $M_j[i] = \max(M^*[i], T_j[i])$, $0 \le i < s$. Applying it to (9), we can express the intersection sketch $M_\cap$ as

$$\begin{aligned} M_\cap[i] &= \min_{j\in[1,t]}\Big\{ \max\big(M^*[i]\,,\ T_j[i]\big)\Big\} \\ &= \max\Big(M^*[i]\,,\ \min_{j\in[1,t]}\big\{\,T_j[i]\,\big\}\Big). \end{aligned} \tag{12}$$

Suppose the value of the $i$th register in $M_\cap$ is $k$, i.e., $M_\cap[i] = k$, $k \ge 0$. There are two cases:

I. The persistent elements in $S^*$ set the value of register $M^*[i]$ to be $k$, and the transient elements in at least one period set the value of $M^*[i]$ no larger than $k$. Namely, $M^*[i] = k$ and $T[i] = \min_{j\in[1,t]}\big\{\,T_j[i]\,\big\} \le k$.

II. The persistent elements in $S^*$ set the value of register $M^*[i]$ smaller than $k$. Of the $t$ HLL sketches of transient elements, the minimum value in this register is exactly $k$. Namely, $M^*[i] < k$ and $T[i] = \min_{j\in[1,t]}\big\{\,T_j[i]\,\big\} = k$.

To calculate the probability for $M_\cap[i] = k$, we should first analyze the probabilistic distribution for $M^*[i]$ and $T[i]$ to carry a particular value. Let $n^*[i]$ be the total number of persistent elements recorded in the $i$th register of the sketch $M^*$, and $n'_j[i]$ be the number of transient elements in the $i$th register of the sketch $T_j$, $j \in [1, t]$. Since each persistent element in $S^*$ randomly selects a register in $M^*$, it has a probability of $\frac{1}{s}$ to map into $M^*[i]$. Hence $n^*[i]$ approximately follows a binomial distribution, $n^*[i] \sim Bino(n^*, \frac{1}{s})$, and the probability for $M^*[i]$ to record $\nu$ persistent elements is

$$\mathbb{P}(n^*[i] = \nu) = \binom{n^*}{\nu}\big(\frac{1}{s}\big)^\nu\big(1 - \frac{1}{s}\big)^{n^*-\nu}. \tag{13}$$

According to the HLL algorithm, the random variable $M^*[i]$ is the maximum value of $\nu$ random variables that are independently and geometrically distributed according to $\mathbb{P}(Y > k) = 2^{-k}, k \ge 0, \nu > 0$. Thus, the cumulative distribution function of $M^*[i]$ under the condition $n^*[i] = \nu, \nu > 0$ is $\mathbb{P}(M^*[i] \le k \,|\, n^*[i] = \nu,\ \nu > 0) = (1 - \frac{1}{2^k})^\nu$. Since $0^0 = 1$ and $\mathbb{P}(M^*[i] \le k \,|\, n^*[i] = \nu,\ \nu = 0) = 1$, the above conditional cumulative distribution function is also satisfied if $\nu = 0$. Combining these two cases, we have

$$\mathbb{P}(M^*[i] \le k \,|\, n^*[i] = \nu) = \big(1 - \frac{1}{2^k}\big)^\nu\,,\ \nu \ge 0\,,\ k \ge 0. \tag{14}$$

Therefore, based on (13) and (14), the cumulative distribution function $F_{M^*[i]}(k)$ of $M^*[i]$ is

$$\begin{aligned} F_{M^*[i]}(k) &= \mathbb{P}(M^*[i] \le k) \\ &= \sum_{\nu=0}^{n^*} \mathbb{P}(M^*[i] \le k \,|\, n^*[i] = \nu) \cdot \mathbb{P}(n^*[i] = \nu) \\ &= \sum_{\nu=0}^{n^*} \binom{n^*}{\nu}\big(\frac{1}{s}\big)^\nu\big(1 - \frac{1}{s}\big)^{n^*-\nu}\big(1 - \frac{1}{2^k}\big)^\nu. \end{aligned} \tag{15}$$

In most situations, the persistent spread $n^*$ is at least 20 and $1/s$ is smaller than or equal to 0.05 ($s \ge 20$). Hence, Poisson distribution can be used to approximate the binomial distribution for efficient calculation, $Bino(n^*, 1/s) \approx Pois(\lambda = \frac{n^*}{s})$. Thereby, we have

$$\begin{aligned} F_{M^*[i]}(k) &\approx \sum_{\nu=0}^{n^*} \frac{\lambda^\nu e^{-\lambda}}{\nu!}\big(1 - \frac{1}{2^k}\big)^\nu = e^{-\lambda}\sum_{\nu=0}^{n^*}\frac{\big(\lambda(1 - \frac{1}{2^k})\big)^\nu}{\nu!} \\ &\approx e^{-\lambda}e^{\lambda(1-\frac{1}{2^k})} = e^{-\frac{\lambda}{2^k}} = e^{-\frac{n^*}{s2^k}}. \end{aligned} \tag{16}$$

Similarly, we calculate the cumulative distribution function $F_{T_j[i]}(k)$

$$\begin{aligned} F_{T_j[i]}(k) &= \sum_{\nu=0}^{n'_j}\binom{n'_j}{\nu}\big(\frac{1}{s}\big)^\nu\big(1 - \frac{1}{s}\big)^{n'_j-\nu}\big(1 - \frac{1}{2^k}\big)^\nu \\ &\approx e^{-\frac{n'_j}{s2^k}} = e^{-\frac{n_j-n^*}{s2^k}}. \end{aligned} \tag{17}$$

As we assume that transient elements in different periods are approximately independent, the cumulative distribution function $F_{T[i]}(k)$ of $T[i]$ is

$$F_{T[i]}(k) = \mathbb{P}(T[i] \le k) = 1 - \mathbb{P}(T[i] > k) \tag{18}$$

$$= 1 - \mathbb{P}\big(\min\{\,T_j[i]\,\}_{j\in[1,t]} > k\big) \approx 1 - \prod_{j=1}^{t}\mathbb{P}(T_j[i] > k)$$

$$= 1 - \prod_{j=1}^{t}\big(1 - \mathbb{P}(T_j[i] \le k)\big) = 1 - \prod_{j=1}^{t}\big(1 - F_{T_j[i]}(k)\big).$$

Therefore, the probability $\mathbb{P}'_i$ for the first case that $M_\cap[i] = k$ is

$$\begin{aligned} \mathbb{P}'_i &= \mathbb{P}(M^*[i] = k) \cdot \mathbb{P}(T[i] \le k) \\ &= \begin{cases} F_{M^*[i]}(k) \cdot F_{T[i]}(k) & k = 0, \\ \big[F_{M^*[i]}(k) - F_{M^*[i]}(k-1)\big] \cdot F_{T[i]}(k) & k \ge 1. \end{cases} \end{aligned}$$

The probability $\mathbb{P}''_i$ for the second case that $M_\cap[i] = k$ is

$$\begin{aligned} \mathbb{P}''_i &= \mathbb{P}(M^*[i] < k) \cdot \mathbb{P}(T[i] = k) \\ &= \begin{cases} 0 & k = 0, \\ F_{M^*[i]}(k-1) \cdot \big[F_{T[i]}(k) - F_{T[i]}(k-1)\big] & k \ge 1. \end{cases} \end{aligned}$$

To sum up, the probability for $M_\cap[i] = k$ is

$$\mathbb{P}(M_\cap[i] = k) = \mathbb{P}'_i + \mathbb{P}''_i \tag{19}$$

$$= \begin{cases} F_{M^*[i]}(k) \cdot F_{T[i]}(k) & k = 0, \\ F_{M^*[i]}(k) \cdot F_{T[i]}(k) - F_{M^*[i]}(k-1) \cdot F_{T[i]}(k-1) & k \ge 1. \end{cases}$$

Let a generation function $G_s(n_1, n_2, \ldots, n_t, n^*, k)$ represent the expression $F_{M^*[i]}(k) \cdot F_{T[i]}(k)$. Combining (16), (17) and (18), then we have

$$\begin{aligned} G_s(n_1, n_2, \ldots, n_t, n^*, k) &= F_{M^*[i]}(k) \cdot F_{T[i]}(k) \\ &\approx e^{-\frac{n^*}{s2^k}} \cdot \Big(1 - \prod_{j=1}^{t}\Big(1 - e^{-\frac{n_j-n^*}{s2^k}}\Big)\Big). \end{aligned} \tag{20}$$

Note that the $n_1, n_2, \ldots, n_t$ can be estimated using the HLL algorithm on HLL sketches $M_1$, $M_2$, . . ., $M_t$, respectively. Thus, they can be treated as constant in the generation function $G$, so we can simplify $G_s(n_1, n_2, \ldots, n_t, n^*, k)$ to $G_s(n^*, k)$. We position $s$ as the subscript of function $G$, rather than its input parameter, because the number of registers in each period is determined by the available memory and is typically a fixed value. Therefore, the probability that the $i$th register in the intersection sketch $M_\cap$ has the value $k$ is

$$\mathbb{P}(M_\cap[i] = k) = \begin{cases} G_s(n^*, k) & k = 0, \\ G_s(n^*, k) - G_s(n^*, k-1) & k \ge 1. \end{cases} \tag{21}$$

In practice, a register can only carry a value in a specific range due to the limited memory size (e.g., 5 bits per register). Let $H$

be the threshold, which is the maximum value (upper bound) that a register's capacity can record. For instance, if the size of a register is 5 bits, its recording range is from 0 to $2^5 = 32$ (exclusive), and the threshold $H$ is 31. Let $h$ be the size of a register, then $H = 2^h - 1$.

Considering the limited register size, we need to modify probability for $M_\cap \geq H$. Assume the register is assigned to $H$ when its value is out of bound. Hence, we have

$$\begin{aligned} \mathbb{P}(M_\cap[i] = H) &= 1 - \sum_{k=0}^{H-1} \mathbb{P}(M^*[i] = k) \\ &= 1 - G_s(n^*, H-1). \end{aligned} \tag{22}$$

Therefore, the probability distribution function for $M_\cap[i]$ to carry a value $k$ in (21) becomes

$$\mathbb{P}(M_\cap[i] = k) = \begin{cases} G_s(n^*, k) & k = 0, \\ G_s(n^*, k) - G_s(n^*, k-1) & 0 < k < H, \\ 1 - G_s(n^*, k-1) & k = H, \\ 0 & k > H. \end{cases} \tag{23}$$

## 4.2 I-HLL Estimator

We provide the I-HLL estimator for persistent spread $n^*$ based on MLE. To establish the likelihood function, we first measure the number of registers among the $s$ registers in $M_\cap$ that carry the value $k$, which is denoted by $N_k$. The reason why we use $N_k$ instead of $k$ as the observing factor is that the observing space size of $N_k$ is equal to $H$, which is far less than $k$'s observing space size $s$. The probability for observing $N_k$ registers in $M_\cap$ carrying the value $k$ is $\mathbb{P}(M_\cap[i] = k)^{N_k}$, assuming these registers are approximately independent. Hence, the combined probability for observing $N_0, N_1, \ldots, N_H$ under the condition that there are $n^*$ elements in the persistent set is

$$\mathbb{P}(N_0, N_1, \ldots, N_H | n^*) = \alpha \prod_{k=0}^{H} \mathbb{P}(M_\cap[i] = k)^{N_k}, \tag{24}$$

where $\alpha$ is a constant that equals $\frac{s!}{N_0! N_1! \ldots N_H!}$. The likelihood function for observing $N_0, N_1, \ldots, N_H$ with respect to $n^*$ is

$$\mathcal{L}(n^* | N_0, N_1, \ldots, N_H) = \alpha \prod_{k=0}^{H} \mathbb{P}(M_\cap[i] = k)^{N_k}. \tag{25}$$

Taking the logarithm on both sizes of the likelihood function, we obtain the log based likelihood function as follows:

$$\ln \mathcal{L} = \ln \alpha + \sum_{k=0}^{H} N_k \ln \mathbb{P}(M_\cap[i] = k). \tag{26}$$

Taking the partial derivative on log based likelihood function with respect to $n^*$, we obtain

$$\begin{aligned} \frac{\partial \ln \mathcal{L}}{\partial n^*} &= \frac{\partial}{\partial n^*}\Big(\ln \alpha + \sum_{k=0}^{H} N_k \ln \mathbb{P}(M_\cap[i] = k)\Big) \\ &= \sum_{k=0}^{H} N_k \frac{\partial \ln \mathbb{P}(M_\cap[i] = k)}{\partial n^*} = \sum_{k=0}^{H} N_k \frac{\frac{\partial}{\partial n^*}\mathbb{P}(M_\cap[i] = k)}{\mathbb{P}(M_\cap[i] = k)}. \end{aligned} \tag{27}$$

The derivative of $\mathbb{P}(M_\cap[i] = k)$ with respect to $n^*$ for an arbitrary value $k \in [0, H]$ is given as follows,

$$\frac{\partial \mathbb{P}(M_\cap[i] = k)}{\partial n^*} = \begin{cases} \frac{\partial}{\partial n^*} G_s(n^*, k) & k = 0, \\ \frac{\partial}{\partial n^*}\big(G_s(n^*, k) - G_s(n^*, k-1)\big) & 0 < k < H, \\ -\frac{\partial}{\partial n^*} G_s(n^*, k-1) & k = H, \end{cases} \tag{28}$$

where the partial derivative of $G_s(n^*, k)$ over $n^*$ is

$$\begin{aligned} &\tfrac{\partial}{\partial n^*} G_s(n^*, k) \approx \tfrac{1}{s2^k} e^{-\frac{n^*}{s2^k}} \cdot \\ &\bigg(\Big(1 + \textstyle\sum_{j=1}^{t} \big(e^{\frac{n_j - n^*}{s2^k}} - 1\big)^{-1}\Big) \cdot \Big(\prod_{j=1}^{t} \big(1 - e^{-\frac{n_j - n^*}{s2^k}}\big)\Big) - 1\bigg). \end{aligned} \tag{29}$$

The calculation of $\frac{\partial}{\partial n^*} G_s(n^*, k)$ is given in the Appendix A.

The maximum likelihood estimation is to find an estimated persistent spread $\hat{n}^*$ that maximizes the log likelihood function $\ln \mathcal{L}$. Therefore, we obtain an estimator for $n^*$:

$$\hat{n}^* = \arg\max_{n^*} \big\{ \ln \mathcal{L} \big\} = \big\{ n^* \mid \frac{\partial}{\partial n^*} \ln \mathcal{L} = 0 \big\}. \tag{30}$$

As a summary, we define a unified function $f_t$ to give a formal I-HLL estimator $\hat{n}^*$ to measure the persistent spread $n^*$ over an arbitrary number $t$ of time periods, which is equivalent to (30).

Definition 1 (I-HLL Persistent Spread Estimator). *For $t \geq 2$, a unified function to estimate the persistent spread of a flow is*

$$\hat{n}^* = f_t\big(s, M_\cap, \{M_j\}_{j \in [1,t]}\big), \tag{31}$$

*where $s$ is the number of registers in each HLL sketch, $M_j$ is the HLL sketch in the $j$th period ($j \in [1, t]$), and $M_\cap$ is the intersection HLL sketch that equals $M_1 \wedge M_2 \wedge \cdots \wedge M_t$.*

## 4.3 Accuracy Analysis

We analyze the relative bias and relative standard error of I-HLL estimator. We denote the value of the $i$th register of $M_\cap$ by a random variable $X_i$, thereby $\mathbb{P}_{X_i}(k) = \mathbb{P}(M_\cap[i] = k)$. Then the expected value and variance of $\frac{\partial \ln \mathbb{P}_{X_i}(k)}{\partial n^*}$ are

$$\begin{aligned} \mu &= E\big(\tfrac{\partial \ln \mathbb{P}_{X_i}(k)}{\partial n^*}\big) = \textstyle\sum_{k=0}^{H} \big(\tfrac{\partial \ln \mathbb{P}_{X_i}(k)}{\partial n^*}\big) \cdot \mathbb{P}_{X_i}(k), \\ \sigma^2 &= Var\big(\tfrac{\partial \ln \mathbb{P}_{X_i}(k)}{\partial n^*}\big) = \textstyle\sum_{k=0}^{H} \big(\tfrac{\partial \ln \mathbb{P}_{X_i}(k)}{\partial n^*}\big)^2 \cdot \mathbb{P}_{X_i}(k) - \mu^2. \end{aligned} \tag{32}$$

Moreover, the new likelihood function for preserving $X_0 = k_0$, $X_1 = k_1, \cdots, X_{s-1} = k_{s-1}$ can be written as

$$\mathcal{L}(n^* | k_1, k_2, \cdots, k_{s-1}) = \prod_{i=0}^{s-1} \mathbb{P}_{X_i}(k). \tag{33}$$

Note that the above likelihood function is only the original likelihood function multiplied by a constant value $\alpha$. Hence, we can still use the notation $\mathcal{L}$ without confusion.

Taking the logarithm of the likelihood function and the derivative with respect to $n^*$, we have

$$\frac{1}{s} E\Big(\big(\frac{\partial \ln \mathcal{L}}{\partial n^*}\big)^2\Big) = \frac{1}{s} E\Big(\big(\sum_{i=0}^{s-1} \frac{\partial \ln \mathbb{P}_{X_i}(k_i)}{\partial n^*}\big)^2\Big). \tag{34}$$

Since $X_0, X_1, \cdots, X_{s-1}$ are roughly independent, assuming $\psi^2 = \frac{(n^* \sigma)^2}{s}$, then we have

$$\begin{aligned} &\tfrac{1}{s} E\Big(\big(\tfrac{\partial \ln \mathcal{L}}{\partial n^*}\big)^2\Big) \approx \tfrac{1}{s} \textstyle\sum_{i=0}^{s-1} E\Big(\big(\tfrac{\partial \ln \mathbb{P}_{X_i}(k_i)}{\partial n^*}\big)^2\Big) \\ &+ \tfrac{1}{s} \textstyle\sum_{i,j \in [0,s)\ i \neq j} E\big(\tfrac{\partial \ln \mathbb{P}_{X_i}(k_i)}{\partial n^*}\big) E\big(\tfrac{\partial \ln \mathbb{P}_{X_j}(k_j)}{\partial n^*}\big) \\ &= E\Big(\big(\tfrac{\partial \ln \mathbb{P}_{X_i}(k_i)}{\partial n^*}\big)^2\Big) + (s-1) E\big(\tfrac{\partial \ln \mathbb{P}_{X_i}(k_i)}{\partial n^*}\big) E\big(\tfrac{\partial \ln \mathbb{P}_{X_j}(k_j)}{\partial n^*}\big) \\ &= \sigma^2 + \mu^2 + (s-1)\mu^2 = s\mu^2 + \sigma^2 = s\mu^2 + s\big(\frac{\psi}{n^*}\big)^2, \end{aligned}$$

where $\mu = 0$ and $\psi^2$ is

$$\begin{aligned} \psi^2 = &\tfrac{(n^*)^2}{s^3} G_s(n^*, 0) + \tfrac{(n^*)^2 G_s^2(n^*, H-1)}{s^3 2^{2(H-1)} (1 - G_s(n^*, H-1))} \\ &+ \textstyle\sum_{k=1}^{H-1} \tfrac{(n^*)^2 \big(G_s(n^*, k) - 2G_s(n^*, k-1)\big)^2}{s^3 2^{2k} (G_s(n^*, k) - G_s(n^*, k-1))} \end{aligned}$$

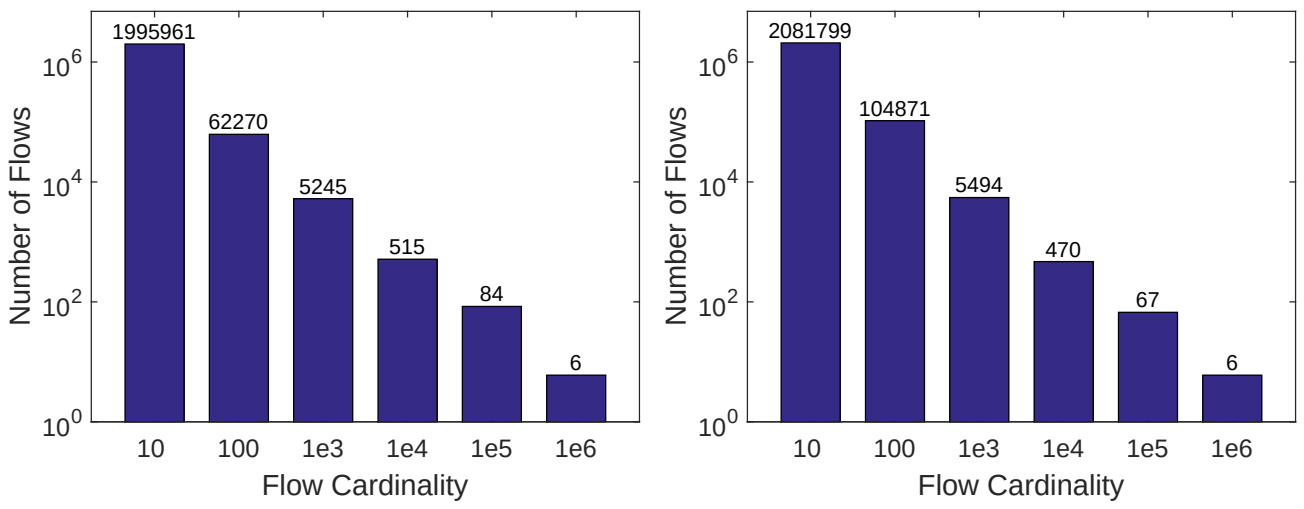

(a) Per-source flow (b) Per-destination flow

Figure 6: Flow cardinality distributions for different flows

The calculation of $\mu$ and $\psi^2$ can be found in the Appendix B.

Hence, the fisher information [25] $I(\hat{n}^*) = \frac{1}{s}E\big((\frac{\partial \ln \mathcal{L}}{\partial n^*})^2\big) = s(\frac{\psi}{n^*})^2$. According to the asymptotic properties of maximum likelihood estimation, our estimator is asymptotically unbiased, and it achieves the Cramer-Rao lower bound:

$$\hat{n}^* \xrightarrow{d} Normal(n^*, \frac{1}{I(\hat{n}^*)}) = Normal(n^*, \frac{(n^*)^2}{s\psi^2}). \tag{35}$$

Therefore, the relative standard error is

$$StdErr(\frac{\hat{n}^*}{n^*}) \approx \frac{1}{\sqrt{s}\psi}, \tag{36}$$

and the $1-\epsilon$ confidence interval for $n^*$ is

$$\hat{n}^* \pm Z_{\frac{\epsilon}{2}} \frac{n^*}{\sqrt{s}\psi}. \tag{37}$$

# 5. VIRTUAL I-HLL ARCHITECTURE

## 5.1 Motivation

In the design of our I-HLL estimator, all flows are allocated with separated and equal-sized HLL sketches to record their elements in each measurement period, which best fits when the flow cardinality is uniformly distributed. However, many studies observe a common fact that the distribution of flow cardinalities is extremely unbalanced in real networks, and small percentage of large flows account for a majority of the Internet traffic (also known as the heavy-tailed distribution). Without loss of generality, we use the real network trace captured by the main gateway of our university as an example. The distributions of per-source flow and per-destination flow are illustrated in Figure 6a and Figure 6b, respectively. Clearly, the vast majority of flows have small cardinalities, while only a small number of flows have large cardinalities. The same trend is observed in the traffic traces from CAIDA [23].

Under this common observation of unbalanced distribution in network traffic data, maintaining one HLL sketch for each flow is not applicable due to the limited size of on-chip SRAM. The reason is that, when we don't know which flows are elephant flows in advance, the sizes of all HLL sketches for I-HLL estimator should be configured according to the largest flow cardinalities in order to achieve reasonably accurate measurement. Therefore, we have to allocate all HLL sketches with the same size that are large enough to accommodate the elephant flows. Hence, for the majority of flows with small cardinalities, the high-order bits in their registers are actually under-utilized as many or even most of them remain zeros, which causes a significant waste of memory. To reduce the memory waste caused by the uneven flow cardinality distribution, register sharing should be enabled among the flows to utilize these unused bits.

## 5.2 Register Sharing and Virtual HLL Sketch

Our idea is to enable register sharing among HLL sketches of all flows. An example is illustrated in Figure 7, where each cell represents a register. The HLL sketches of all flows are no longer separated. Instead, they share registers from a common register pool, called physical register array $A$. Each flow pseudo-randomly picks a number of registers from the physical register array $A$ to form its logical data structure called virtual HLL sketch. Since virtual HLL sketches of all flows share the same register pool $A$, elephant flows can 'borrow' memory from small flows to utilize the unused space.

Table 1: Notations

| | |
|---|---|
| $A$ | a physical array of registers |
| $A_j$ | a physical register array of period $j$ |
| $m$ | number of registers in physical register array |
| $A_{dst}$ | virtual HLL sketch of flow $dst$ |
| $s$ | number of registers used by virtual HLL sketch |
| $H_i(dst)$ | hash function that maps the $i$th register of $A_{dst}$ to $A$ |
| $n^*$ | number of persistent elements of flow $dst$ |
| $\hat{n}^*$ | an estimation of $n^*$ |
| $n_s^*$ | number of persistent elements in $A_{dst}$ |
| $\hat{n}_s^*$ | an estimation of $n_s^*$ |
| $n_u^*$ | number of persistent elements in $A$ |
| $\hat{n}_u^*$ | an estimation of $n_u^*$ |

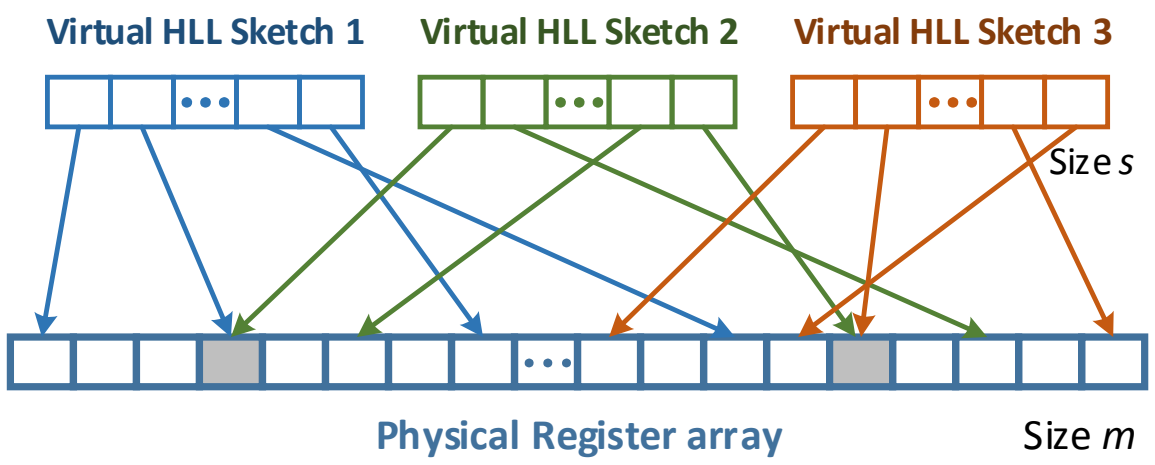

Figure 7: Register sharing and virtual HLL sketch.

From above, we design a novel persistent spread estimation architecture based on virtual HLL sketches on top of register sharing, called Virtual Intersection HyperLogLog estimator (VI-HLL), where each flow is allocated with a virtual HLL sketch of multiple registers in each measurement period. Suppose the total memory size of $A$ is $M$ bits, and the size of each register is $h$ bits. So the number of registers in $A$ is $m = \frac{M}{h}$. Each virtual HLL sketch is configured a unified size $s$ that is large enough to accommodate all flows. For each flow $dst$, we randomly select $s$ registers from $A$ to form its virtual HLL sketch $A_{dst}$. The $i$th register in $A_{dst}$, denoted by $A_{dst}[i]$, can be selected from $A$ as follows,

$$A_{dst}[i] = A[H_i(dst)], \quad 0 \le i < s, \tag{38}$$

where $H_i()$ is a hash function whose range is $[0, m)$. The hash function $H_i()$ can be implemented using one master hash function $H$,

$$H_i(dst) = H(dst \oplus R[i]), \quad 0 \le i < s, \tag{39}$$

where $H()$ is a hash function whose range is $[0, m)$, $\oplus$ is the XOR operator, and $R$ is an array of $s$ random seeds.

In the next subsections, we will introduce our VI-HLL architecture to estimate the persistent spread simultaneously for multiple flows. The architecture includes two components; one for recording flow elements in $A$, and the other for estimating the persistent spread for an arbitrary flow $dst$. The frequently used notations are summarized in Table 1 for quick reference.

## 5.3 Record Flow Elements in $A$

In each time period, a register array $A$ of $m$ registers is used to record elements information of all flows. At the beginning of each period, all registers of $A$ are initialized to zeros. In technical discussion below, we again consider per-destination flows through a router that measures the distinct number of source addresses in each flow. When a packet arrives, the router extracts its flow label $dst$ and treats the source address $src$ as an element of flow $dst$. The router records the element in the flow's virtual HLL sketch $A_{dst}$. To do so, it first performs a hash $H(src)$, whose binary representation is denoted as $x$. Let $p$ is the leading $b$ ($b = \log_2 s$) bits in $x$, and $q$ is the remaining bits:

$$p = \langle x_1 x_2 \ldots x_b \rangle,$$
$$q = \langle x_{b+1} x_{b+2} \ldots \rangle.$$

Using the value of $p$, the router maps the element $src$ of flow $dst$ pseudorandomly to a register of its virtual HLL sketch $A_{dst}[p]$, and updates the value $A_{dst}[p]$ if its current value is smaller than $\rho(q)$,

$$A_{dst}[p] = \max\big(A_{dst}[p], \rho(q)\big). \tag{40}$$

Applying (38) and (39), we have

$$A[H(dst \oplus R[p])] = \max\big(A[H(dst \oplus R[p])], \rho(q)\big). \tag{41}$$

The online recording module for one time period is summarized in Algorithm 2. At the end of each measurement period, the physical register array $A$ will be offloaded from on-chip SRAM to main memory of a server for long-term storage and offline query. Assume that we have measured $t$ consecutive time periods, thereby we have $t$ physical register arrays, which are denoted by $A_1, A_2, \ldots, A_t$.

**Algorithm 2** Online recording module for one time period

1: Initialize a register array $A$ of size $m$ with all zeros;
2: **for** package $\langle src, dst \rangle$ **do**
3: $\quad x := H(src);\quad p := \langle x_1 x_2 \ldots x_b \rangle;\quad q := \langle x_{b+1} x_{b+2} \ldots \rangle;$
4: $\quad i := H(dst \oplus R[p]);\quad A[i] = \max\big(A[i], \rho(q)\big);$
5: **end for**
6: **return** $A$ at the end of the measurement period

## 5.4 VI-HLL Estimator

We describe our VI-HLL estimator, which uses the sequence of physical register arrays $A_1, A_2, \ldots, A_t$, to estimate the persistent spread for an arbitrary flow.

Consider a flow $dst$ under query, we reconstruct its virtual HLL sketch $M$ from an arbitrary physical register array $A$, where the $i$th register in virtual HLL sketch has been mapped to the register $A[H_i(dst)]$ in $A$, $M[i] = A_{dst}[i] = A[H_i(dst)]$, $0 \le i < s$. So

$$M = \big\langle A_{dst}[0], A_{dst}[1], \ldots, A_{dst}[s-1] \big\rangle$$
$$= \big\langle A[H_0(dst)], A[H_2(dst)], \ldots, A[H_{s-1}(dst)] \big\rangle.$$

Since we have $t$ physical register arrays $A_1, A_2, \ldots, A_t$, we can reconstruct $t$ virtual HLL sketches, denoted as $M_1, M_2, \ldots, M_t$. Then we have the virtual intersection HLL sketch $M_\cap$: $M_\cap = M_1 \wedge M_2 \wedge \cdots \wedge M_t$. An intuitive method is that we can apply our previous I-HLL estimator on $M_1, M_2, \ldots, M_t$ and $M_\cap$ in Definition 1, to filter the transient elements and estimate the cardinality of persistent elements in the virtual HLL sketch.

However, this intuitive method will cause overestimating problem for the persistent spread $n^*$ of flow $dst$. This is because the virtual HLL sketch of flow $dst$ not only records the persistent elements belonging to flow $dst$, but also contains the persistent elements coming from other flows due to the mechanism of register sharing. Specifically, if some of registers shared with other flows happen to be set by some persistent elements of these flows, then these registers will be updated in all virtual HLL sketches $M_1, M_2, \ldots, M_t$ in all $t$ periods such that they are recorded in $M_\cap$. Therefore, the persistent elements introduced by register sharing, called noise, causes the overestimation problem when estimating the persistent spread of the flow $dst$ with I-HLL estimator.

Our VI-HLL estimator is to remove the noise that comes from other flows, and gives unbiased persistent spread estimation. Let $n^*$ be the number of persistent elements of flow $dst$, $n_s^*$ be the number of persistent elements recorded in virtual intersection HLL sketch $M_\cap$ of flow $dst$, and $n_u^*$ be the number of persistent elements in physical intersection register array $A_\cap = A_1 \wedge A_2 \wedge \cdots \wedge A_t$. Due to register sharing, we know that $n_s^*$ is the persistent spread $n^*$ of flow $dst$ plus the noise (persistent spreads) introduced by other flows. Let $Y$ be a random variable for the number of noise persistent spreads recorded by the virtual intersection HLL sketch $M_\cap$, then we have

$$Y = n_s^* - n^*. \tag{42}$$

To recover $n^*$ from virtual HLL sketches of flow $dst$, we remove such noise as follows. The total number of persistent elements coming from other flows is $n_u^* - n^*$. From the view of the flow $dst$, these elements are noise. As we assume that there are many flows and $n^* \ll n_u^*$, each noise element from other flows has approximately the same probability to map into $M_\cap$. This probability is equal to $\frac{s}{m}$ due to the random selection of $s$ registers by the virtual HLL sketch from $A$ ($m$ registers). Hence, $Y$ follows a binomial distribution, $Y \sim Bino(n_s^* - n^*, \frac{s}{m})$. The expected number of noise elements mapped to $M_\cap$ is $E(Y) = \frac{s(n_u^* - n^*)}{m}$. Therefore, we have

$$E(n_s^* - n^*) = E(Y) = \frac{s(n_u^* - n^*)}{m}. \tag{43}$$

By the law of large numbers in probability theory, if the number of $s$ is large, the relative variance $Var(\frac{n_s^* - n^*}{E(n_s^* - n^*)})$ approaches to zero. In this case, the expected value $E(n_s^* - n^*)$ can be approximated by an instance value, $n_s^* - n^*$. Hence, we have

$$n_s^* - n^* \approx \frac{s(n_u^* - n^*)}{m} \;\Rightarrow\; n^* \approx \frac{ms}{m-s}\Big(\frac{n_s^*}{s} - \frac{n_u^*}{m}\Big). \tag{44}$$

Based on Definition 1, we can obtain accurate estimation $\hat{n}_s^*$ and $\hat{n}_u^*$ over $M_\cap$ and $A_\cap$, respectively.

$$\hat{n}_s^* = f_t\big(s, M_\cap, \{M_j\}_{j\in[1,t]}\big),$$
$$\hat{n}_u^* = f_t\big(m, A_\cap, \{A_j\}_{j\in[1,t]}\big).$$

Therefore, we obtain the estimate for persistent spread $n^*$:

$$\hat{n}^* = \frac{ms}{m-s}\Big(\frac{\hat{n}_s^*}{s} - \frac{\hat{n}_u^*}{m}\Big). \tag{45}$$

The VI-HLL estimator for flow $dst$ is summarized in Algorithm 3.

**Algorithm 3** VI-HLL persistent spread estimator for flow $dst$

1: **Input:** $s$, $m$, $\{M_j\}_{j\in[1,t]}$ and $\{A_j\}_{j\in[1,t]}$.
2: **Step 1:** Obtain the virtual intersection HLL sketch of flow $dst$:
$\quad M_\cap \leftarrow M_1 \wedge M_2 \wedge \cdots \wedge M_t$.
$\quad$ Estimate $n_s^*$ in $M_\cap$ by Definition 1:
$\quad \hat{n}_s^* := f_t\big(s, M_\cap, \{M_j\}_{j\in[1,t]}\big)$.
3: **Step 2:** Obtain the intersection HLL sketch of all flows:
$\quad A_\cap \leftarrow A_1 \wedge A_2 \wedge \cdots \wedge A_t$.
$\quad$ Estimate $n_u^*$ in $A_\cap$ by Definition 1:
$\quad \hat{n}_u^* := f_t\big(m, A_\cap, \{A_j\}_{j\in[1,t]}\big)$.
4: **Step 3:** Remove noise and obtain estimation of $n^*$ by (45):
$\quad \hat{n}^* := \frac{ms}{m-s}\big(\frac{\hat{n}_s^*}{s} - \frac{\hat{n}_u^*}{m}\big)$.
5: **return** the estimated persistent spread $\hat{n}^*$.

## 5.5 Accuracy Analysis

We now analyze the relative bias and relative standard error of our VI-HLL estimator. According to analysis of I-HLL estimator in Section 4.3, we have the following theorem.

THEOREM 1. *Let $n_s^*$ be the number of persistent elements that are mapped to the virtual HLL sketch. Suppose the number $s$ of registers is large enough. Then,*

$$E(\hat{n}_s^*) \approx n_s^*$$
$$Var(\hat{n}_s^*) \approx \frac{(n_s^*)^2}{s\psi_s^2} \tag{46}$$
$$StdErr\big(\frac{\hat{n}_s^*}{n_s^*}\big) \approx \frac{1}{\sqrt{s}\psi_s}$$

*where $\psi_s$ is a variance related to $s$ and $n_s^*$.*

### 5.5.1 Relative Bias

According to (42), we know that $n_s^* = n^* + Y$, and $Y$ follows a binomial distribution of $Binom(n_u^* - n^*, \frac{s}{m})$. Combing Theorem 1, under the condition of $Y = l$, $l \in [0, n_u^* - n^*]$, we have

$$E(\hat{n}_s^* \,|\, Y = l) \approx n_s^* = n^* + l, \tag{47}$$

and

$$\mathbb{P}(Y = l) = \binom{n_u^* - n^*}{l}\big(\frac{s}{m}\big)^l\big(1 - \frac{s}{m}\big)^{n_u^* - n^* - l}. \tag{48}$$

By (47) and (48), we can calculate

$$\begin{aligned} E(\hat{n}_s^*) &= \sum_{l=0}^{n_u^*-n^*} E(\hat{n}_s^* \,|\, Y = l) \cdot \mathbb{P}(Y = l) \\ &\approx \textstyle\sum_{l=0}^{n_u^*-n^*} (n^* + l) \cdot \binom{n_u^* - n^*}{l}\big(\frac{s}{m}\big)^l\big(1 - \frac{s}{m}\big)^{n_u^* - n^* - l} \\ &= n^* + E(Y) = n^* + \frac{s(n_u^* - n^*)}{m}. \end{aligned} \tag{49}$$

The value of $\hat{n}_u^*$ is estimated based on the physical register array $A$. Therefore, $E(\hat{n}_u^*) \approx n_u^*$. From the definition in (1) and estimation formula (45), the relative bias of $\hat{n}^*$ is

$$\begin{aligned} Bias\big(\frac{\hat{n}^*}{n^*}\big) &= E\big(\frac{\hat{n}^*}{n^*}\big) - 1 = \frac{ms}{m-s}\Big(\frac{E(\hat{n}_s^*)}{sn^*} - \frac{E(\hat{n}_u^*)}{mn^*}\Big) - 1 \\ &\approx \frac{ms}{m-s}\Big(\frac{n^* + \frac{s(n_u^*-n^*)}{m}}{sn^*} - \frac{\hat{n}_u^*}{mn^*}\Big) - 1 = 0. \end{aligned} \tag{50}$$

Hence, the VI-HLL estimator $\hat{n}^*$ is approximately unbiased for $n^*$.

### 5.5.2 Relative Standard Error

Next we derive the relative standard error of $\hat{n}^*$. Under the condition of $Y = l$, by Theorem 1, we have

$$Var(\hat{n}_s^* \,|\, Y = l) \approx \frac{(n^* + l)^2}{s\psi_s^2} \tag{51}$$

Similarly, we have

$$Var(\hat{n}_u^*) \approx \frac{(n_u^*)^2}{m\psi_m^2}, \tag{52}$$

where $m$ is the number of registers in $A$. By Theorem 1 and (47),

$$\begin{aligned} &E\big((\hat{n}_s^*)^2 \,|\, Y = l\big) = Var(\hat{n}_s^* \,|\, Y = l) + E\big(\hat{n}_s^* \,|\, Y = l\big)^2 \\ &\approx \frac{(n^* + l)^2}{s\psi_s^2} + (n^* + l)^2 = \big(1 + \frac{1}{s\psi_s^2}\big)(n^* + l)^2. \end{aligned} \tag{53}$$

Combining (48) and above equation, we have

$$\begin{aligned} E\big((\hat{n}_s^*)^2\big) &= \sum_{l=0}^{n_u^*-n^*} E\big((\hat{n}_s^*)^2 | Y = l\big) \cdot \mathbb{P}(Y = l) \\ &\approx \textstyle\sum_{l=0}^{n_u^*-n^*} \big(1 + \frac{1}{s\psi_s^2}\big)(n^* + l)^2 \cdot \binom{n_u^* - n^*}{l}\big(\frac{s}{m}\big)^l\big(1 - \frac{s}{m}\big)^{n_u^* - n^* - l} \\ &= \big(1 + \frac{1}{s\psi_s^2}\big)\Big(\big(n^* + \frac{s(n_u^*-n^*)}{m}\big)^2 + \frac{s(n_u^*-n^*)}{m}\big(1 - \frac{s}{m}\big)\Big). \end{aligned} \tag{54}$$

Hence, the variance of the estimation $n^*$ is

$$\begin{aligned} Var(\hat{n}^*) &= \big(\frac{ms}{m-s}\big)^2\Big(\frac{Var(\hat{n}_s^*)}{s^2} - \frac{Var(\hat{n}_u^*)}{m^2}\Big) \\ &\approx \big(\frac{m}{m-s}\big)^2\Big(\frac{1}{s\psi_s^2}\big(n^* + \frac{s(n_u^*-n^*)}{m}\big)^2 \\ &\quad + \big(\frac{1}{s\psi_s^2} + 1\big)\frac{s(n_u^*-n^*)}{m}\big(1 - \frac{s}{m}\big) - \big(\frac{s}{m}\big)^2\frac{(n_u^*)^2}{m\psi_m^2}\Big). \end{aligned} \tag{55}$$

The relative standard error of $n^*$ is

$$\begin{aligned} StdErr(\hat{n}^*) &= \frac{\sqrt{Var(\hat{n}^*)}}{n^*} \approx \frac{m}{(m-s)n^*} \cdot \\ &\sqrt{\frac{1}{s\psi_s^2}\big(n^* + \frac{s(n_u^*-n^*)}{m}\big)^2 + \big(\frac{1}{s\psi_s^2} + 1\big)\frac{s(n_u^*-n^*)}{m}\big(1 - \frac{s}{m}\big) - \big(\frac{s}{m}\big)^2\frac{(n_u^*)^2}{m\psi_m^2}}. \end{aligned} \tag{56}$$

# 6. SIMULATIONS

In this section, we use extensive simulations to evaluate our persistent spread estimator VI-HLL based on register sharing. We compare it with state-of-the-art V-Bitmap [22]. Since our goal is to design persistent spread estimator that can be used in tight memory space while delivering high accuracy, in our simulations, we only consider memory requirements that are less than 3 bits per flow. We also evaluate the impact of the number of periods $t$, signal-to-noise ratio $SNR_j$, and number of registers $s$ on the VI-HLL performance.

## 6.1 Simulation Setup

We implement VI-HLL as well as V-Bitmap, and compare them through extensive simulations. The data we used is simulated from real-world network traffic traces. Note that the flows in the simulations can be per-source flows, per-destination flows, or other user-defined flows, which all lead to similar results. Without loss of generality, we use per-destination flows for presentation. The traffic data in each period contains $124,846,736$ distinct elements generated by $11,453,043$ flows. The average flow cardinality is $10.90$ per flow. We simulate tremendous users concurrently accessing a large server farm, which is quite practical in today's main gateway router. Some of the flow elements are persistent elements, which exist throughout the $t$ periods, and the rest are transient elements. In each period, we control the ratio of persistent elements to the transient elements by signal-to-noise ratio $SNR_j = \frac{n^*}{n_j'} = \frac{n^*}{n_j - n^*}$, $j \in [1, t]$.

The performance metrics used in our simulations include memory requirement and estimation accuracy as discussed in Section 2. We run two sets of simulations. The first set is used to evaluate the impact of memory size on the persistent spread estimation accuracy of V-Bitmap and VI-HLL. We vary the memory space $M$ from 0.5MB, 1MB, 2MB to 4MB, which translates to approximately 0.37bits/flow, 0.75bits/flow, 1.5bits/flow and 3bits/flow, respectively. To make a fair comparison, VI-HLL and V-Bitmap are given the same memory size to process the simulated traffic data in each case. For V-Bitmap, the length of each virtual bitmap is configured as $10,000$ to achieve better accuracy, or as large as $50,000$ to accommodate the large flows to have larger estimation range as in [22]. For VI-HLL, we use a virtual HLL sketch to record each flow in each period. The length $s$ of each virtual HLL sketch is configured as $512$. The second set of simulations evaluates the impact of different parameters on the performance of VI-HLL. We fix the memory size to $M = $ 2MB, and vary $t$, $SNR_j$ and $s$ with different values to observe their impact on estimation accuracy. The simulation results are given as follows.

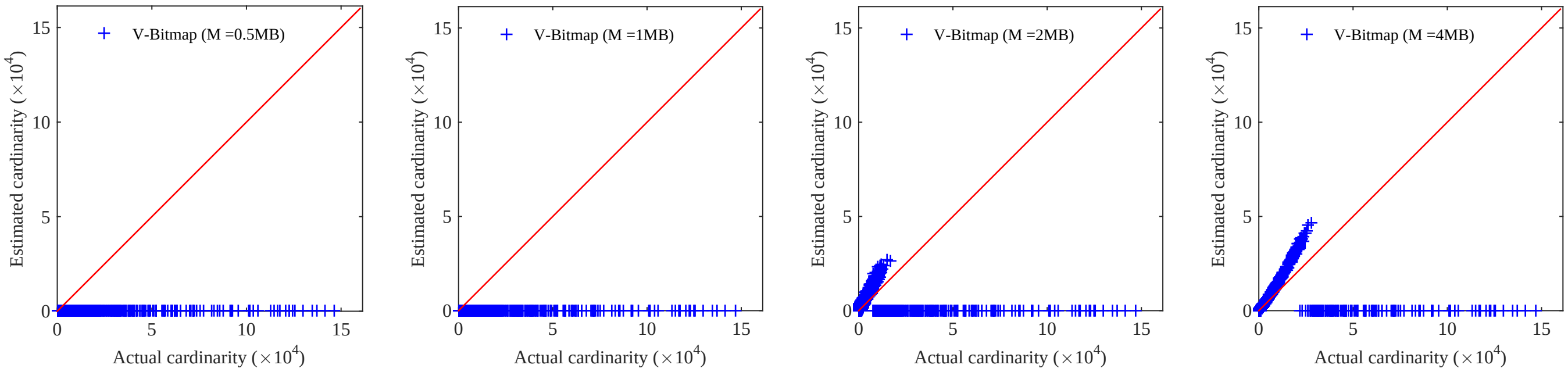

(a) V-Bitmap with 0.37 bit per flow (b) V-Bitmap with 0.75 bit per flow (c) V-Bitmap with 1.5 bits per flow (d) V-Bitmap with 3 bits per flow

Figure 8: Persistent spread estimation using V-Bitmap under different memory overhead, with $t = 10$, $SNR_j = 1$ and $s = 10000$.

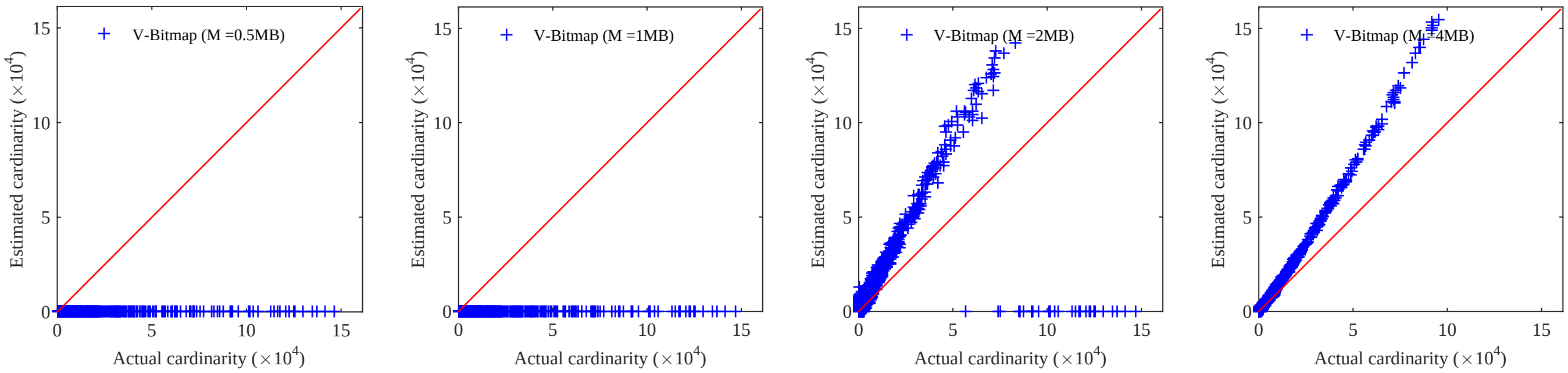

(a) V-Bitmap with 0.37 bit per flow (b) V-Bitmap with 0.75 bit per flow (c) V-Bitmap with 1.5 bits per flow (d) V-Bitmap with 3 bits per flow

Figure 9: Persistent spread estimation using V-Bitmap under different memory overhead, with $t = 10$, $SNR_j = 1$ and $s = 50000$.

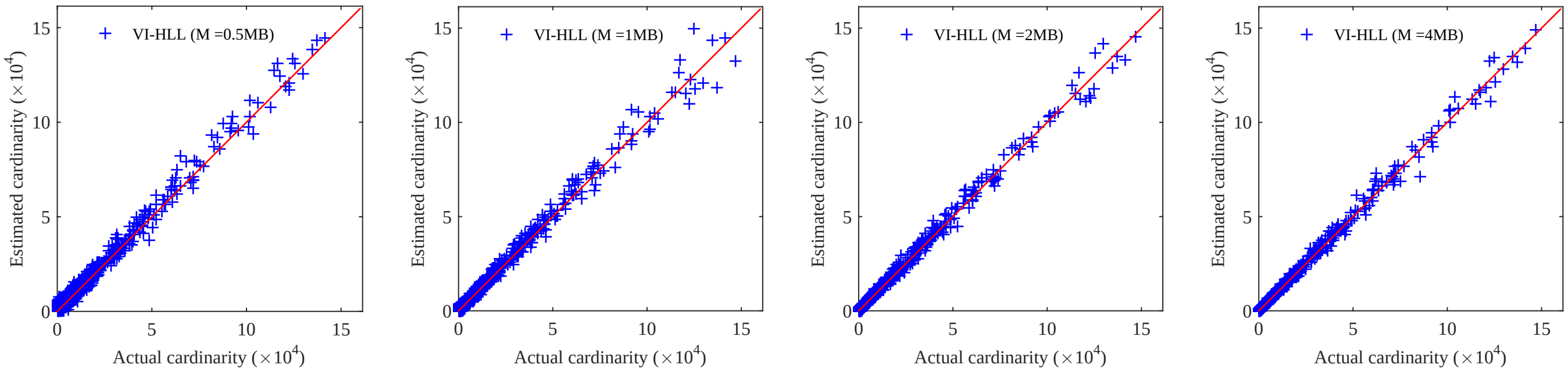

(a) VI-HLL with 0.37 bit per flow (b) VI-HLL with 0.75 bit per flow (c) VI-HLL with 1.5 bits per flow (d) VI-HLL with 3 bits per flow

Figure 10: Persistent spread estimation using VI-HLL under different memory overhead $M$, with $t = 10$, $SNR_j = 1$ and $s = 512$.

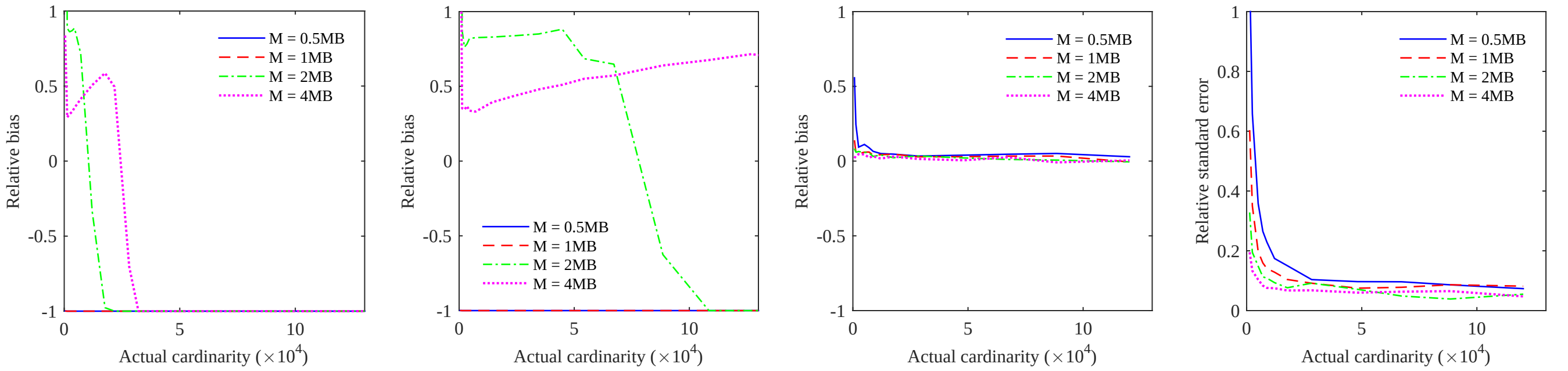

(a) V-Bitmap estimation bias (s = 10000) (b) V-Bitmap estimation bias (s = 50000) (c) V-IHLL estimation bias (s = 512) (d) V-IHLL estimation accuracy (s = 512)

Figure 11: Compare VI-HLL and V-Bitmap under different memory overhead $M$.

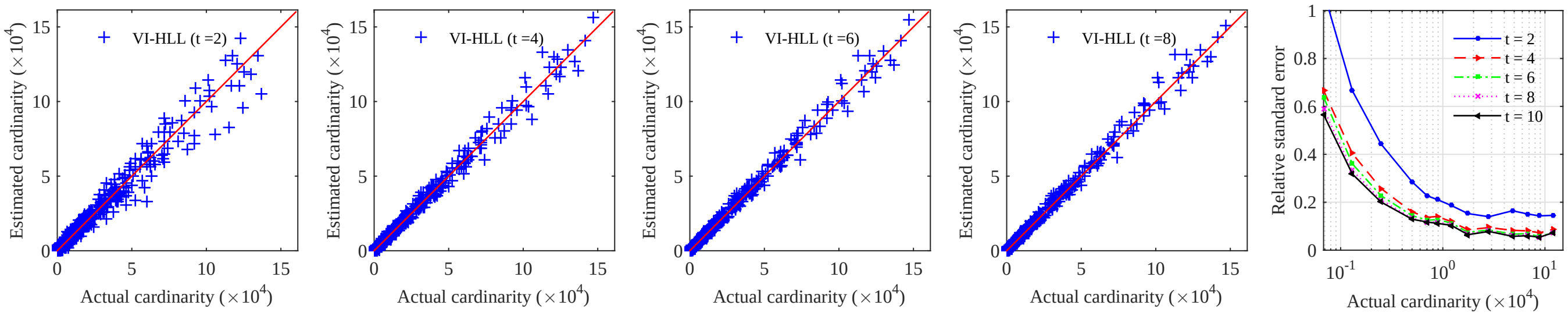

Figure 12: Estimation results and relative errors of VI-HLL under different values of $t$, with $M$ = 2MB, $SNR_j$ = 1 and $s$ = 512.

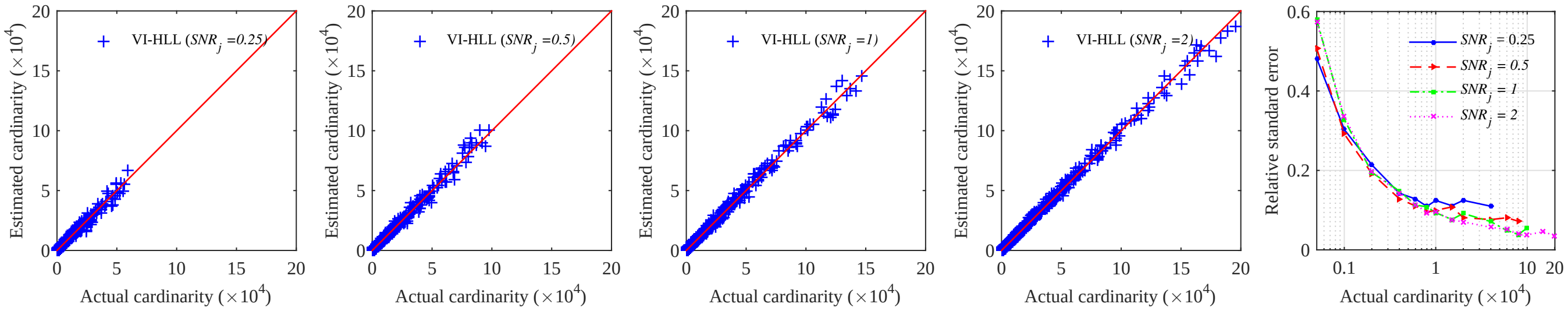

Figure 13: Estimation results and relative errors of VI-HLL under different values of $SNR_j$, with $M$ = 2MB, $t$ = 10 and $s$ = 512.

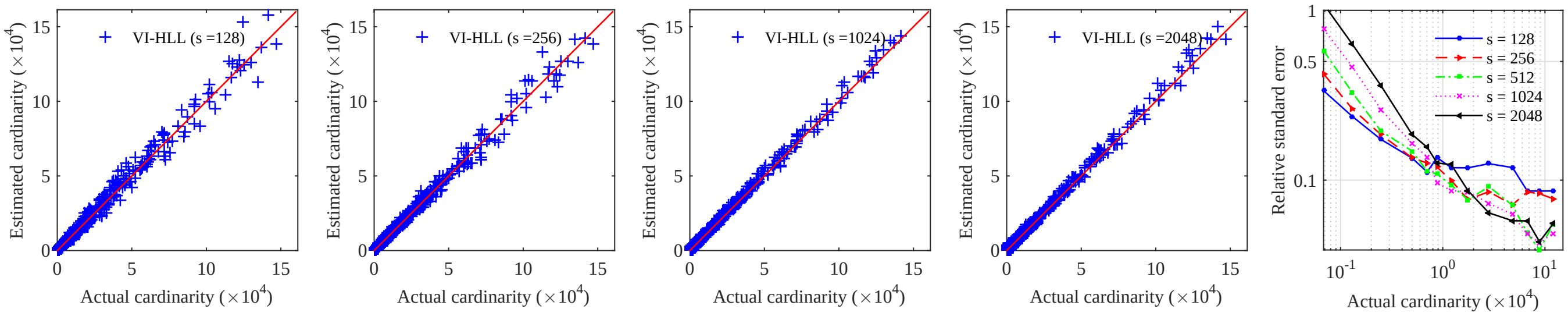

Figure 14: Estimation results and relative errors of VI-HLL under different values of $s$, with $M$ = 2MB, $t$ = 10 and $SNR_j$ = 1 .

## 6.2 VI-HLL v.s. V-Bitmap

We study the estimation accuracy of V-Bitmap and VI-HLL with the available memory ranging from 0.5MB, 1MB, 2MB to 4MB. The total measurement time periods $t$ is fixed to 10, and signal-to-noise ratio $SNR_j$ is 1. The comparison results of V-Bitmap and VI-HLL are presented in Figures 8, 9, 10 and 11. The first three figures show the estimation results for V-Bitmap with $s = 10000$, V-Bitmap with $s = 50000$, and VI-HLL with $s = 512$, each of which includes four plots under different memory sizes $M$. Each point in each plot represents a flow, where the $x$ coordinate is the actual persistent spread cardinality $n^*$ and the $y$ coordinate is the estimated cardinality $\hat{n}^*$. The equality line, $y = x$, is also shown. Clearly, the closer a point is to the equality line, the more accurate the estimate is.

In Figure 8, plot(a) and plot(b) show when the available memory is tight, e.g., $M$ = 0.5MB (0.37bits/flow) or $M$ = 1MB (0.75bits/flow), V-Bitmap (s = 10000) cannot give reasonable estimations for most flows. The reason is that the estimation accuracy of V-Bitmap depends on the fill rate – the proportion of bits in a bitmap that are set to be one. The higher the fill rate, the worse the accuracy. For example, when $M$ = 0.5MB, each bit is mapped by 30 elements on average, so almost all bits are set to 1. Hence, V-Bitmap can no longer work in such tight memory. As the memory size increases from 1MB to 4MB as shown in plot(c) and plot(d), V-Bitmap generates some positively biased results, but still cannot yield estimates for large persistent spread flows due to the high fill rate. Although increasing $M$ can enlarge the estimation range of V-Bitmap to some extent, it still does not address the problem caused by high fill rate.

An alternative way to extend the estimation range for V-Bitmap is to increase the virtual bitmap size $s$. Figure 9 gives the simulations results for V-Bitmap with $s = 50000$. Clearly, it still cannot work under tight memory as shown in plot(a) and plot(b). When increasing memory size as illustrated in plot(c) and plot(d), V-Bitmap gives larger estimation range comparing with the last two plots in Figure 8, but the results are still quite inaccurate because the fill rate is still very high and large size bitmap introduces more noise. Therefore, V-Bitmap cannot work under tight memory.

Figure 10 shows the simulation results of VI-HLL when $s = 512$. Clearly, VI-HLL can generate very accurate persistent spread estimates for both small and large flows as points are clustered to the equality line for all four plots. This is true even under a tight memory, e.g., $M$ = 0.5MB (0.37bit/flow) as shown in plot(a). In addition, through register intersection, VI-HLL can easily handle wide estimation ranges without modifying preset parameters, which is required by V-Bitmap in order to generate sound measurement results when facing different traffic situations. VI-HLL provides a more robust solution for real-life persistent spread measurement.

The relative bias $Bias(\frac{\hat{n}^*}{n^*})$ of V-Bitmap and VI-HLL and relative standard error $StdErr(\frac{\hat{n}^*}{n^*})$ of VI-HLL are given in Figure 11. Plot(a), plot(b) and plot(c) present the estimation bias of V-Bitmap with $s = 10000$, V-Bitmap with $s = 50000$ and VI-HLL, respectively. We can see that under tight memory, V-Bitmap has large bias, while VI-HLL has small relative bias and relative standard errors. Also, VI-HLL becomes more accurate when more memory is used.

## 6.3 Impact of Value $t$ on VI-HLL

In our second set of simulations, we firstly study the impact of the number of time periods $t$ on the performance of VI-HLL. We fix $M$ = 2MB, $SNR_j = 1$ and $s = 512$, and vary $t$ from 10 to 2, 4, 6 to 8. The results are presented in Figure 12. The first four plots are estimation results under $t = 2$, 4 , 6 and 8. Corresponding relative standard errors are illustrated in the fifth plot. Clearly, when $t$ becomes larger, the relative standard error becomes smaller, which reflects an interesting feature of VI-HLL that its estimation accuracy improves when the number of time periods increases. This is because VI-HLL detects the existence of the persistent elements from the register intersection on all HLL sketches $M_1, M_2, \ldots, M_t$. The probability for an intersection register in $M_\cap$ to be updated higher by transient elements, captured by the term $\mathbb{P}_i''$ in (19), decreases as $t$ value grows. Therefore, VI-HLL permits network admin to set arbitrarily large $t$ values to differentiate persistent and transient elements.

## 6.4 Impact of Value $SNR_j$ on VI-HLL

Next, we evaluate the impact of the signal-to-noise ratio $SNR_j$ on the performance of VI-HLL. We fix $M$ = 2MB, $t = 10$ and $s = 512$, and vary $SNR_j$ from 0.25, 0.5, 1 to 2. The results are presented in Figure 13. The first four plots are estimation results under $SNR_j$ = 0.25, 0.5, 1 and 2. Corresponding relative standard errors are illustrated in the fifth plot. From the plots, we see that the accuracy degrades a bit as $SNR_j$ decreases, but VI-HLL still renders reasonably high accuracy. The ability of tolerating heavy noise in VI-HLL makes it more flexible to use in practice.

## 6.5 Impact of Value $s$ on VI-HLL

Finally, we investigate the impact of the register size $s$ on the performance of VI-HLL. We fix $M$ = 2MB, and vary the value of $s$ from 512 to 128, 256, 1024 to 2048. The results are represented in Figure 14. The first four plots are estimation results under $s = 128$, 256, 1024 and 2048. Corresponding relative standard errors are illustrated in the fifth plot. Clearly, when $s$ is relatively small ($s = 128$), the relative standard errors are larger than when $s = 256$ or $s = 512$ for large size flows. However, when $s$ gets large enough ($s = 1024$ or $s = 2048$), the estimation accuracy for large size flows stabilizes, but the estimation accuracy for small size flows becomes noticeably worse. Combining these two effects, in practice, it may be more appropriate to choose a virtual HLL sketch size of either 512 or 1024.

# 7. CONCLUSION

In this paper, we propose a highly compact and efficient Virtual Intersection HyperLogLog (VI-HLL) architecture for persistent spread measurement. It can help detect long-term stealthy network activities in the background of short-term activities of legitimate users. Through extensive analysis and simulations, we demonstrate that VI-HLL can perform well even in a very tight memory space (less than 3 bits or even 0.37 bits per flow) with wide measurement range and reasonably high accuracy. Therefore, it can be implemented in fast on-chip SRAM to keep up with the line speed of modern routers, or low-cost commodity computers to process big network data.

# 8. ACKNOWLEDGMENT

This work is supported in part by the National Science Foundation under grant STC-1562485 and a grant from Florida Center for Cybersecurity.

# APPENDIX

## A. THE PARTIAL DERIVATIVE OF $G_s(n^*, k)$

We have shown the generating function $G_s(n^*, k)$ in (20). Now we analyze its partial derivative over the persistent spread $n^*$.

Firstly, we analyze the partial derivatives of the two main components of $G_s(n^*, k)$, i.e., $e^{-\frac{n^*}{s2^k}}$ and $\prod_{j=1}^{t}\left(1-e^{-\frac{n_j-n^*}{s2^k}}\right)$,

$$\frac{\partial}{\partial n^*}e^{-\frac{n^*}{s2^k}} = -\frac{1}{s2^k}e^{-\frac{n^*}{s2^k}},$$

$$\frac{\partial}{\partial n^*}\prod_{j=1}^{t}\left(1-e^{-\frac{n_j-n^*}{s2^k}}\right)$$

$$=\sum_{i=1}^{t}\left(\frac{-\frac{1}{s2^k}e^{-\frac{n_i-n^*}{s2^k}}}{1-e^{-\frac{n_i-n^*}{s2^k}}}\cdot\prod_{j=1}^{t}\left(1-e^{-\frac{n_j-n^*}{s2^k}}\right)\right)$$

$$=\left(\sum_{j=1}^{t}\frac{-\frac{1}{s2^k}}{e^{\frac{n_j-n^*}{s2^k}}-1}\right)\cdot\left(\prod_{j=1}^{t}\left(1-e^{-\frac{n_j-n^*}{s2^k}}\right)\right).$$

Then, we apply the above equation and have

$$\frac{\partial}{\partial n^*}G_s(n^*,k) \approx -\frac{1}{s2^k}e^{-\frac{n^*}{s2^k}}\cdot\left(1-\prod_{j=1}^{t}\left(1-e^{-\frac{n_j-n^*}{s2^k}}\right)\right)$$
$$+\frac{1}{s2^k}e^{-\frac{n^*}{s2^k}}\cdot\left(\sum_{j=1}^{t}\left(e^{\frac{n_j-n^*}{s2^k}}-1\right)^{-1}\right)\cdot\left(\prod_{j=1}^{t}\left(1-e^{-\frac{n_j-n^*}{s2^k}}\right)\right).$$

Finally, we can simplify the above equation as

$$\frac{\partial}{\partial n^*}G_s(n^*,k) \approx \frac{1}{s2^k}e^{-\frac{n^*}{s2^k}}\cdot$$
$$\left(\left(1+\sum_{j=1}^{t}\left(e^{\frac{n_j-n^*}{s2^k}}-1\right)^{-1}\right)\cdot\left(\prod_{j=1}^{t}\left(1-e^{-\frac{n_j-n^*}{s2^k}}\right)\right)-1\right).$$

## B. ANALYSIS OF $\mu$, $\sigma^2$ AND $\psi^2$

Next, we analyze $\mu$ and $\sigma^2$ based on their definitions in (32). Applying (28), we have

$$\mu = E\left(\frac{\partial \ln \mathbb{P}_{X_i}(k)}{\partial n^*}\right) = \sum_{k=0}^{H}\left(\frac{\partial \ln \mathbb{P}_{X_i}(k)}{\partial n^*}\right)\cdot\mathbb{P}_{X_i}(k)$$
$$=\sum_{k=0}^{H}\left(\frac{\frac{\partial \mathbb{P}_{X_i}(k)}{\partial n^*}}{\mathbb{P}_{X_i}(k)}\right)\cdot\mathbb{P}_{X_i}(k) = \sum_{k=0}^{H}\frac{\partial \mathbb{P}_{X_i}(k)}{\partial n^*}$$
$$=\frac{\partial}{\partial n^*}\sum_{k=0}^{H}\mathbb{P}_{X_i}(k) = \frac{\partial 1}{\partial n^*} = 0,$$

and

$$\sigma^2 = Var\left(\frac{\partial \ln \mathbb{P}_{X_i}(k)}{\partial n^*}\right) = \sum_{k=0}^{H}\left(\frac{\partial \ln \mathbb{P}_{X_i}(k)}{\partial n^*}\right)^2\cdot\mathbb{P}_{X_i}(k).$$
$$=\sum_{k=0}^{H}\left(\frac{\frac{\partial \mathbb{P}_{X_i}(k)}{\partial n^*}}{\mathbb{P}_{X_i}(k)}\right)^2\cdot\mathbb{P}_{X_i}(k) = \sum_{k=0}^{H}\frac{\left(\frac{\partial \mathbb{P}_{X_i}(k)}{\partial n^*}\right)^2}{\mathbb{P}_{X_i}(k)}$$
$$=\frac{\left(\frac{\partial G_s(n^*,0)}{\partial n^*}\right)^2}{G_s(n^*,0)}+\frac{\left(\frac{\partial G_s(n^*,H-1)}{\partial n^*}\right)^2}{1-G_s(n^*,H-1)}$$
$$+\sum_{k=1}^{H-1}\frac{\left(\frac{\partial}{\partial n^*}(G_s(n^*,k)-G_s(n^*,k-1))\right)^2}{G_s(n^*,k)-G_s(n^*,k-1)},$$

where

$$G_s(n^*,k) \approx e^{-\frac{n^*}{s2^k}}\cdot\left(1-\prod_{j=1}^{t}\left(1-e^{-\frac{n_j-n^*}{s2^k}}\right)\right)$$
$$\frac{\partial}{\partial n^*}G_s(n^*,k) \approx \frac{1}{s2^k}e^{-\frac{n^*}{s2^k}}\cdot$$
$$\left(\left(1+\sum_{j=1}^{t}\left(e^{\frac{n_j-n^*}{s2^k}}-1\right)^{-1}\right)\cdot\left(\prod_{j=1}^{t}\left(1-e^{-\frac{n_j-n^*}{s2^k}}\right)\right)-1\right).$$

Assuming that the signal $n^*$ is independent with the noise $n_j - n*$ (i.e., $\frac{\partial n'_j}{\partial n^*} = 0$), then

$$\frac{\partial}{\partial n^*}G_s(n^*,k) \approx -\frac{1}{s2^k}e^{-\frac{n^*}{s2^k}}\cdot\left(1-\prod_{j=1}^{t}\left(1-e^{-\frac{n_j-n^*}{s2^k}}\right)\right)$$
$$= -\frac{1}{s2^k}G_s(n^*,k).$$

Therefore, we can simplify the $\sigma^2$ as

$$\sigma^2=\frac{1}{s^2}G_s(n^*,0)+\frac{G_s^2(n^*,H-1)}{s^2 2^{2(H-1)}(1-G_s(n^*,H-1))}$$
$$+\sum_{k=1}^{H-1}\frac{\left(G_s(n^*,k)-2G_s(n^*,k-1)\right)^2}{s^2 2^{2k}(G_s(n^*,k)-G_s(n^*,k-1))}.$$

Since $\psi^2 = \frac{(n^*\sigma)^2}{s}$, we have

$$\psi^2=\frac{(n^*)^2}{s^3}G_s(n^*,0)+\frac{(n^*)^2 G_s^2(n^*,H-1)}{s^3 2^{2(H-1)}(1-G_s(n^*,H-1))}$$
$$+\sum_{k=1}^{H-1}\frac{(n^*)^2\left(G_s(n^*,k)-2G_s(n^*,k-1)\right)^2}{s^3 2^{2k}(G_s(n^*,k)-G_s(n^*,k-1))}.$$